\documentclass[aps,pre,citeautoscript,onecolumn,superscriptaddress,nofootinbib,nobibnotes,10pt]{revtex4-1}
\usepackage{graphicx}
\usepackage{dcolumn}
\usepackage{bm}
\usepackage{amsmath,amssymb}
\usepackage[utf8]{inputenc}
\usepackage[T1]{fontenc}
\usepackage{mathptmx}
\usepackage{etoolbox}
\usepackage{lineno}
\makeatletter
\def\@email#1#2{%
 \endgroup
 \patchcmd{\titleblock@produce}
  {\frontmatter@RRAPformat}
  {\frontmatter@RRAPformat{\produce@RRAP{*#1\href{mailto:#2}{#2}}}\frontmatter@RRAPformat}
  {}{}
}%
\makeatother

\begin{document}

\title{Aberration stabilized light bullets in normal dispersion injected Kerr resonators}
\author{Yannick Fleck}
\affiliation{Institute for Theoretical Physics, University of Münster, Wilhelm-Klemm-Str. 9, 48149 Münster,
Germany}%
\author{Thomas G. Seidel}%
\affiliation{Institute for Theoretical Physics, University of Münster, Wilhelm-Klemm-Str. 9, 48149 Münster, Germany}%

\author{Jesús Yelo-Sarrión}
\affiliation{Departament de Física, Universitat de les Illes Balears, \& Institute of Applied Computing and Community Code (IAC-3),
C/ Valldemossa km 7.5, 07122 Mallorca, Spain
}%

\author{Julien Javaloyes}
\affiliation{Departament de Física, Universitat de les Illes Balears, \& Institute of Applied Computing and Community Code (IAC-3),
C/ Valldemossa km 7.5, 07122 Mallorca, Spain
}%

\author{Svetlana V. Gurevich}
\affiliation{Institute for Theoretical Physics, University of Münster, Wilhelm-Klemm-Str. 9, 48149 Münster,
Germany}%
\affiliation{
Center for Data Science and Complexity (CDSC), University of Münster, Corrensstr. 2, 48149 Münster, Germany}

\date{\today}

\begin{abstract}
We investigate three-dimensional light bullets (LBs) in an optically injected, vertically emitting Kerr–Gires–Tournois interferometer operating in the normal dispersion regime. In the uniform field limit, the dynamics is governed by a generalized Lugiato–Lefever equation with third-order dispersion, paraxial diffraction, and fourth-order diffraction, stemming from lens and mirror aberrations close to the self-imaging condition. Using direct numerical simulations, we show that without aberration only transversely drifting, tilted multi-peak localized states (LSs) are stable in two dimensions, whereas no stable LBs exist in three dimensions. However, a small negative spherical aberration stabilizes a rich multistable family of two-dimensional LSs, including straight and tilted multi-peak LSs, oscillating states, and complex bound states, which persist even for negative paraxial diffraction. We identify an Andronov–Hopf bifurcation followed by a homoclinic transition toward tilted states, accompanied by hysteresis. Finally, we demonstrate that stable straight and tilted LBs exist in three dimensions in the presence of abberation, the latter forming a continuous family parametrized by the transverse tilt direction.
\end{abstract}

\maketitle

\section{Introduction}

The understanding of self-organized spatiotemporal patterns is key in nonlinear optics and the formation of multidimensional localized structures has been the subject of substantial interest over recent decades~\cite{SFA-PRL-00,BTB-NAT-02,PLG-PRL-08,AFO-AAM-09,GRMc-PRL-17,IKC-NatPhot-23,KAM-NatRevPhys-19}, to cite a few. In particular, the observation of stable~\emph{light bullets (LBs)}, consisting of pulses of light that are simultaneously confined in the transverse and propagation directions and preserve their shape in the course of their propagation remains one of the long-standing challenges in photonics.

Indeed, the optical confinement scenario that would lead to LBs is envisioned through conservative mechanisms in which a self-focusing nonlinearity compensates for the spreading effect of chromatic dispersion and/or diffraction. However, in~\cite{S-OL-90} was demonstrated that conservative Kerr LBs found in the nonlinear Schrödinger equation are unstable and collapse~\cite{Z-SJETP-72,RR-PhScr-86} in three dimensions.

During the last years, other confinement mechanisms were envisioned in different nonlinear dissipative optical systems including parabolic modification of the transverse
refractive index profile of the material, bistable cavities with an instantaneous response of the active medium, saturable absorption, photonic lattices, nonlocal and quadratic nonlinearities and, recently, in passive cavities, to name just a few~\cite{YCL-OptCom-95,TM-PRL-99,VVK-OS-00,BKW-PRE-02,BMP-PRL-04,CPM-NJP-06,KAM-NatRevPhys-19,GPT-PRL-21}.
Robust dissipative three-dimensional and high-order solitons and
breathers were predicted to exist in passively driven nonlinear cavities in the presence of an attractive three-dimensional parabolic potential~\cite{SPRM-PRL-23,SPRM-CSF-24}. Further, stable LBs were predicted in the output of a wide aperture passively mode-locked laser operated in the long cavity regime~\cite{J_PRL_16,GJ-PRA-17,PJG-PTRA-18,DJG-Chaos-20}, whereas discrete LBs were theoretically predicted in an array of passively mode-locked lasers coupled
via evanescent fields in a ring geometry~\cite{SPJG-Chaos-20}.

Another confinement mechanism relies on the use of high-order dispersive effects. In particular, small negative fourth-order dispersion has shown to arrest spatiotemporal collapse of LBs with anomalous dispersion in a planar waveguide with pure Kerr nonlinearity~\cite{FI-OL-04}. Further, the formation of LBs in waveguides with a parabolic refractive index profile and pure quartic chromatic dispersion was reported recently in~\cite{PRSM-PRA-24}. In general, high-order dispersive effects has not only proved to be a key for stabilizing localized states and frequency combs in optical systems~\cite{TG-OL-10,PGL-OL-14,PGG_PRA_17,TABB-PRA-20} and the compensation of chromatic dispersion extended the level of control upon pattern
formation in the temporal domain, leading not only to
to the observation of purely quartic solitons~\cite{KH-OC-94,BSS-NAC-16,DRH-APLP-21}, but also allowing e.g., the control of modulational instabilities~\cite{TML-OL-07,DKB-OL-13}.

In the spatial domain, higher order spatial operators naturally appear in e.g., large aspect ratio lasers \cite{Lega_1994}, fiber lasers~\cite{KZC-PhotRes-21}, photorefractive and semiconductor materials~\cite{HMT-PRL-95,SSW-PRL-97,MM-PRE-02} as well as in metamaterials~\cite{GCdST-PRA-07,KTV-JOSAB-09,TKG-PRA-11}, to name just a few. Recently, it was shown that the interplay between optical aberrations and the proximity to the self-imaging condition, allows to cancel or reverse paraxial diffraction in nearly degenerate broad-area surface-emitting lasers~\cite{GMJ-PRR-24}, giving rise to temporally localized Turing patterns, spatially tilted beams or off-axis temporal localized states~\cite{BVM-Optica-22,BVM-PR-23}. Close to the self-imaging condition, paraxial diffraction can be strongly reduced, and the leading transverse contribution then stems from the spherical aberration of the imaging optics, which acts as a fourth-order diffraction operator. It can therefore be seen as the spatial counterpart of fourth-order dispersion. One may thus expect it to provide a confinement mechanism in the transverse plane that arrests collapse, in the same way as fourth-order dispersion does in the temporal domain~\cite{FI-OL-04}.

In this paper, we discuss how Kerr LBs appear in an optically injected, diffractive broad area vertical-emission Kerr–Gires–Tournois interferometer~\cite{SPV-OL-19,SJG-OL-22,KSG-OL-22,SKJGW_Chaos_23,SJG-OL-24,KGJ-CSF-26,HSJ-Chaos-26} operating in the normal dispersion regime in the presence of fourth order diffraction stemming from the lens and mirror aberrations close to the self-imaging condition~\cite{GMJ-PRR-24}. We shall show that, the fourth-order diffraction induced by the aberration provides the transverse length scale and is the key ingredient that stabilizes multidimensional LSs and LBs. In particular, in the absence of aberrations, the additional transverse dimension strongly restricts the set of stable solutions. In two dimensions, only tilted two- and three-peak localized states (LSs) are stable, and they drift in both the longitudinal and transverse directions. In three dimensions, no stable LBs could be found. We show that introducing a small negative spherical aberration acts as an efficient stabilizing mechanism. It gives rise to a rich multistable set of straight, tilted and oscillating multi-peak LSs, as well as complex bound states. These persist over wide ranges of injection and diffraction, even for negative paraxial diffraction. By following these branches, we find evidence of a supercritical Andronov–Hopf bifurcation leading to transversely wiggling states. This is followed by a homoclinic transition towards tilted states. Finally, we demonstrate that stable straight and tilted LBs exist in three dimensions. The tilted LBs form a continuous family parametrized by the orientation of the tilt in the transverse plane.

The paper is organized as follows. In Section 2, we introduce the generalized spatio-temporal Lugiato–Lefever model describing the system. Section 3 is devoted to the dynamics of LSs in two dimensions and to the stabilizing role of the spherical aberration. In Section 4, we extend the analysis to three-dimensional LBs. Finally, Section 5 summarizes our results and discusses perspectives.

\section{Model System}

\begin{figure}[h!]
 \centering
 \includegraphics[width=0.5\columnwidth]{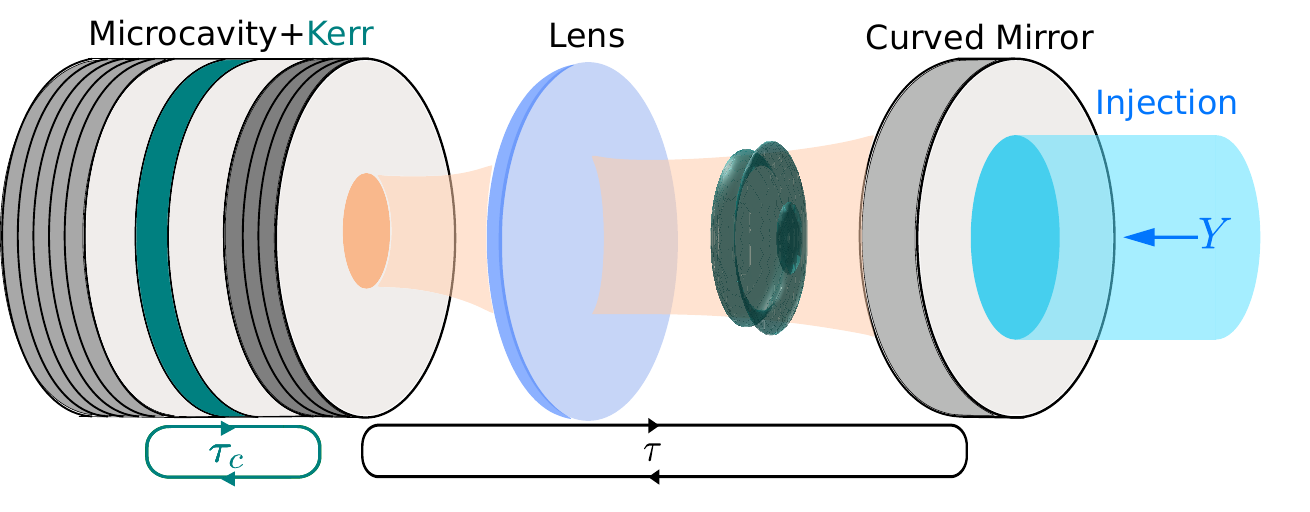}
 \caption{Schematic of a micro-cavity containing a thin slice of Kerr material coupled face to face to a distant external mirror by an imperfect self-imaging system. It is driven by a monomode injection field with amplitude $Y$.
 }\label{fig:setup}
\end{figure}

The KGTI system is composed of a monomode microdisk cavity of a few micrometers containing a Kerr medium, see Fig~\ref{fig:setup}. The micro-cavity of the round-trip $\tau_c$ is closed by two distributed Bragg mirrors and is coupled to a long external cavity of a few centimeters with round-trip time $\tau$ typically a few centimeters long which is closed by a mirror with reflectivity $\eta$ and the phase $\phi$. This whole system is subjected to continuous-wave injection with amplitude $Y_0$ and frequency $\omega_0$.

In~\cite{SPV-OL-19,SJG-OL-22}, a theoretical model for the KGTI system in a temporal domain relying on delay algebraic equations was introduced (see the full model derivation from the first principles in~\cite{KGJ-CSF-26}). There, in the normal dispersion regime, dark and bright temporal localized states (LSs) were found to form via connecting fronts between bistable continuous wave (CW) homogeneous background states. The latter interlock at multiple equilibrium distances leading to a rich ensemble of multistable solutions. The obtained LSs are typically in the picosecond range, thereby resulting in optical frequency combs having a THz bandwidth and tunable GHz repetition rate. Futhermore, it was demonstrated~\cite{SGJ-PRL-22} that for small cavity losses and weak injection, i.e. in the so-called uniform field limit, the temporal dynamics of the KGTI system can be well-described by the Lugiato-Lefever Equation (LLE)~\cite{LL-PRL-87,HTW-OptCom-92} with third-order dispersion.

In the spatial, transverse domain, we follow the methods developed in~\cite{GMJ-PRR-24} and deduce the spatial operator from first principle calculations using the imaging conditions and employing the Huygens-Fresnel transform~\cite{SIEGMAN-BOOK} for quadratic systems with mirror and lenses described by Gauss ABCD matrices. Here, the effect of lens and mirror aberrations close to the self-imaging condition is derived using perturbative theory and the diffraction in the micro-cavity is considered in the paraxial approximation, see~\cite{GMJ-PRR-24} for details. Our final generalized normalized spatio-temporal LLE model for the field $E=E(x,y,z,t)$ reads:

\begin{equation}
		\partial_{t}E=Y-(1+i\theta)\,E+i\,E\,|E|{{}^2}+i\,d_{2}\,\partial_{z}^2E+d_{3}\partial_{z}^3E+i\,B\,\Delta_{\perp}\,E+i\,S\,\Delta_{\perp}^2E, \label{eq:LLE_norm}
	\end{equation}
Here, intensity is scaled to the Kerr intensity, $\Delta_{\perp}=\partial_x^2+\partial_y^2$ is a transverse Laplacian, $\theta$ and $Y$ are the normalized frequency detuning of the cavity and the amplitude of the driving field, $d_2$ is a group velocity dispersion (GVD) coefficient with (+) for anomalous and (-) for the normal GVD regimes, and $d_3$ is the third order dispersion (TOD) coefficient. The equation is set to the self-focusing regime. Further, $B$ and $S$ correspond to the normalized paraxial diffraction parameter, and the spherical aberration parameter, respectively. In particular, for a transverse plane wave, the transverse operator in Eq.~\eqref{eq:LLE_norm} contributes a phase rate $-i\,(B\,k^2 - S\,k^4)$. Hence, for $S < 0$, the aberration term increases the effective diffraction at large transverse wavenumbers, which counteracts self-focusing at small scales and prevents the formation of arbitrarily narrow structures.


In what follows, unless stated otherwise, we will focus on the study of the generalized LLE~\eqref{eq:LLE_norm}  considering normal GVD regime, i.e. we set $d_2<0$. In this case, as was mentioned above, in the one-dimensional (longitudinal) case, dark stationary LSs form via the locking of fronts that connect the two stable CW states, forming a collapsed snaking bifurcation structure. The fronts can interlock at discrete distances defined by their oscillatory tails. The solution accumulate at the so-called Maxwell line where both fronts exhibit the same drift. The presence of the TOD $d_3\neq0$ enables the formation of bright moving LSs that undergo the same form of collapsed snaking towards the Maxwell point as their intensity increases. With increasing TOD, the bifurcation structure for both bright and dark LSs shifts to higher values of the injection $Y$, and the pinning regions for these structures broaden~\cite{PGL-OL-14,PGG_PRA_17,SPV-OL-19,SJG-OL-22}.

\section{Dynamics of localized states in two dimensions}

We start our analysis with the case of two dimensions, i.e $E=E(x,z,t)$ and $\Delta_{\perp}=\partial_x^2$. To study the impact of the the spherical aberration on the stability of localized states in the presence of the transverse dynamics, we solve the 2D generalized LLE~\eqref{eq:LLE_norm} numerically on the rectangular domain $(x,\,z)\in \Omega=[-30,\,30]\times[-50,\,50]$ with periodic boundary conditions employing a split-step Fourier method on $256\times 256$ grid subjected to noise of a small amplitude.

\begin{figure}[h!]
 \centering
 \includegraphics[width=0.45\columnwidth]{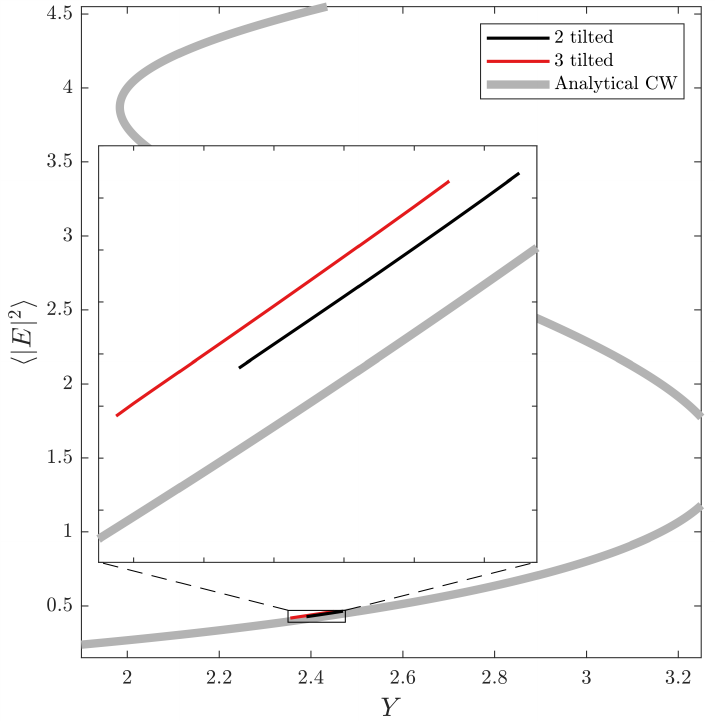}~\includegraphics[width=0.45\columnwidth]{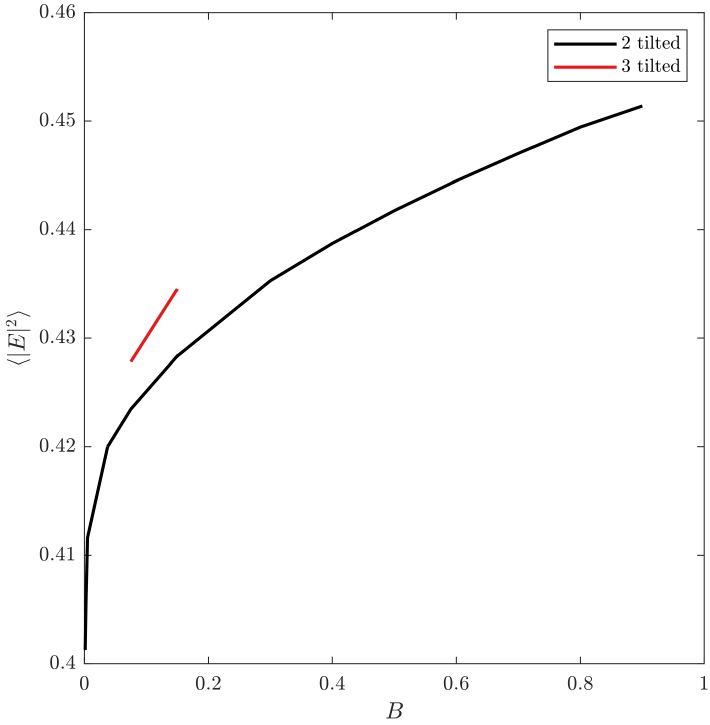}
 \caption{Numerical bifurcation diagram obtained via direct time simulations of the 2D Eq.~\eqref{eq:LLE_norm}. The average intensity of the stable tilted two-peak (black) and three-peak (red) LSs found in the normal dispersion regime for $S=0$ are shown as a function of (left) injection $Y$ at $B=0.15$ and (right) diffraction $B$ at $Y=2.39$. The gray line corresponds to the analytically calculated CW branch, where the stability information is omitted. Other parameters are: $\theta=4$, $d_2=-0.4$, $d_3=0.3$.
 }\label{fig:2dS0_YB}
\end{figure}

\begin{figure}[h!]
 \centering
 \includegraphics[width=0.4\columnwidth]{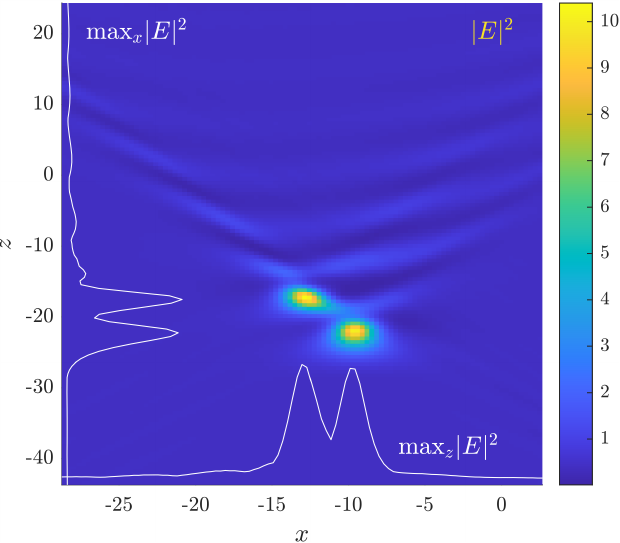}~\includegraphics[width=0.4\columnwidth]{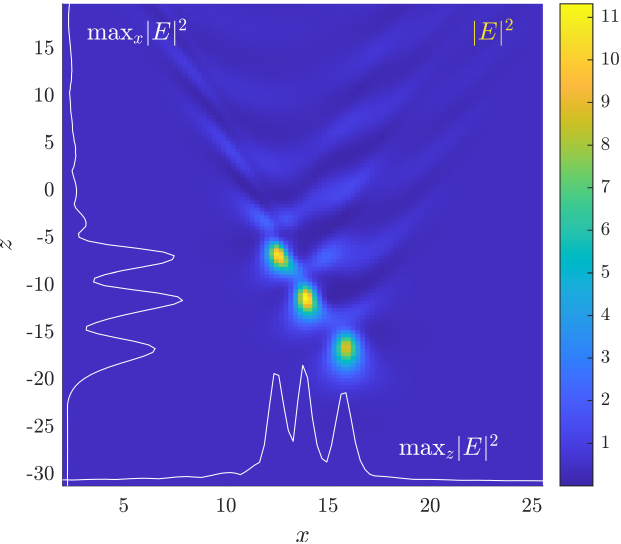}
 \caption{Exemplary two-dimensional tilted two-peak (left) and three-peak (right) LSs obtained in the numerical integration of Eq.~\eqref{eq:LLE_norm} in the normal dispersion regime at $Y=2.39$ for $B=0.8$ (left) and $B=0.15$ (right) (cf. Fig.~\ref{fig:2dS0_YB}). A zoom into the spatial intensity distribution $|E(x,z)|^2$ is shown together with projections in $x$ and $z$ onto the respective axis in white, indicating the intensity maximum over the corresponding axis. Other parameters as in Fig.~\ref{fig:2dS0_YB}.
 }\label{fig:2dS0_B_profiles}
\end{figure}

In order to see the impact of the transversal dimension on the stability of LSs, we first set $S=0$ and perform a numerical scan in the injection $Y$ for a fixed diffraction $B$, keeping other parameters fixed, see Fig.~\ref{fig:2dS0_YB} (left).
Here, only tilted LSs consisting of two (black) and three (red) peaks were found, cf. Fig~\ref{fig:2dS0_B_profiles} for exemplary profiles where a zoom into the spatial intensity distribution $|E(x,z)|^2$ is shown together with projections in $x$ and $z$ onto the respective axis in white, indicating the intensity maximum over the corresponding axis. One note that LSs exhibit oscillatory tails in $z$-direction, due to
the nonzero TOD coefficient. Despite extensive searches, however, no
stable LSs with one or more than four peaks were found, potentially due to their narrow range of existence or instability in the spatio-temporal two-dimensional regime.
Except for the expected uniform drift in the longitudinal $z$-direction due to the presence of TOD, the found LSs also drift in the transverse $x$-direction. Note that the drift direction is arbitrary and both right- and left-tilted LSs can be obtained.  Although we lack information on the unstable states, the bifurcation diagram in Fig.~\ref{fig:2dS0_YB} suggests the presence of possible saddle-node bifurcations at both branch ends. This structure resembles that of the temporal, one-dimensional case, indicating a collapsed snaking structure~\cite{PGL-OL-14,PGG_PRA_17,SPV-OL-19,SJG-OL-22}.
Next, we fixed the injection to the value, where both found LSs exist and performed a numerical scan in the diffraction $B$, see Fig.~\ref{fig:2dS0_YB} (right). One can see that both the field intensity and the transversal velocity decrease with $B$ due to the transversal compression of the LSs profiles. Note that both scans presented in Fig.~\ref{fig:2dS0_YB} revealed the same solution morphology and, without aberration, the transverse dimension strongly restricts the set of stable solutions: only tilted, transversely drifting two- and three-peak LSs survive, in a narrow range of injection, whereas single-peak, straight and dark LSs are absent.

\begin{figure}[h!]
 \centering
 \includegraphics[width=0.44\columnwidth]{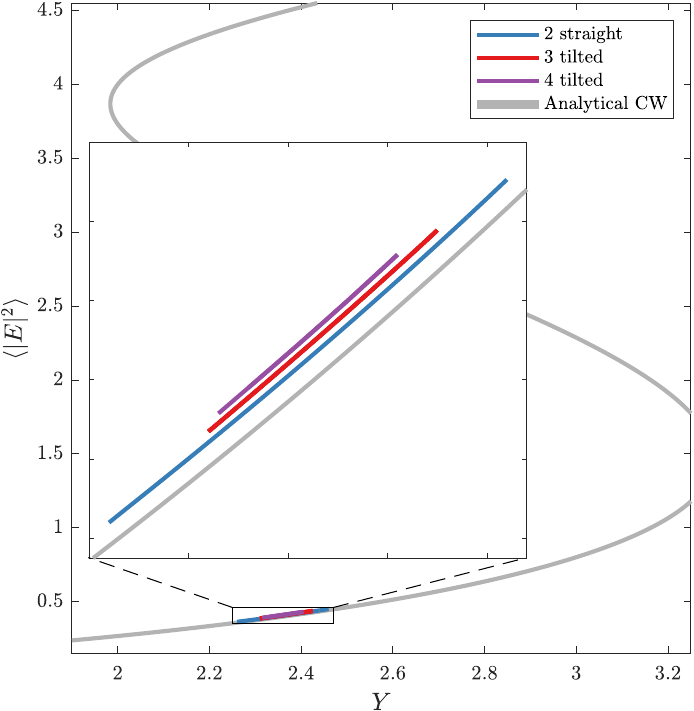}~\includegraphics[width=0.455\columnwidth]{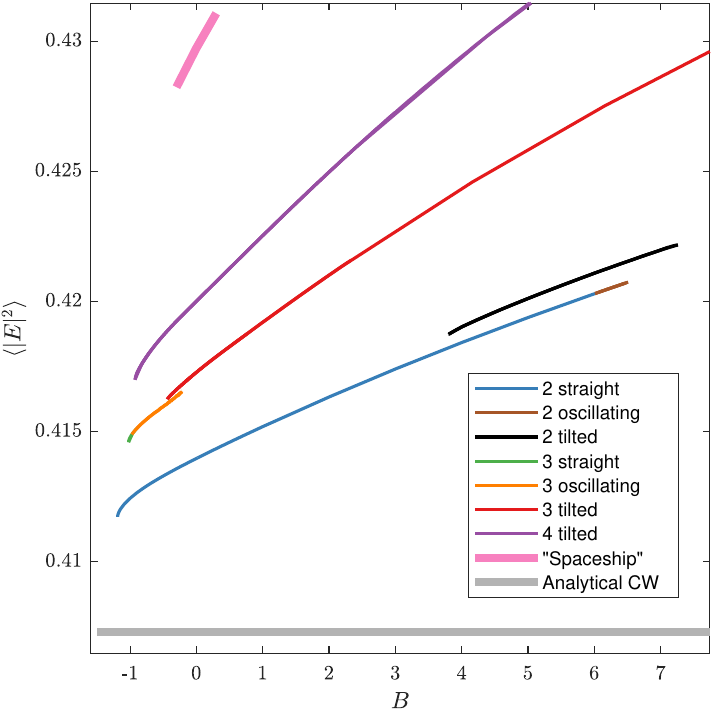}
 \caption{Bifurcation diagram obtained by numerical integration of a 2D Eq.~\eqref{eq:LLE_norm} for $S=-0.5$. The average field intensity as a function of (left) injection $Y$ and (right) diffraction $B$ calculated at $B=0.01$ (left) and $Y=2.38$ (right) is presented. Blue and green lines correspond to an straight two- and three- peak LSs, black, red and violet lines represent tilted two- three- and four-peak LSs, respectively. Two- and three- peak oscillating LSs are depicted in brown and orange, respectively. The gray line corresponds to the analytically calculated CW branch, where the stability information is omitted. Other parameters as in Fig.~\ref{fig:2dS0_YB}.
 }\label{fig:2dSn_Y}
\end{figure}

To study the impact of the aberration parameter $S$ on the stability of the LSs, we set $S$ to a small negative value and performed a scan in the injection $Y$ for fixed small positive diffraction $B=10^{-2}$ where paraxial diffraction alone is weak and only two-peak tilted LSs were stable (cf. Fig.~\ref{fig:2dS0_YB} (right)). The results are presented in Fig.~\ref{fig:2dSn_Y}~(left).
One can see that for $S\neq0$, different co-existing LSs were found to be stable in a large range of the injection. They include straight, collinear LSs consisting of two peaks (blue) as well as tilted LSs, consisting of three- and four-peaks, cf. red (violet) lines in Fig.~\ref{fig:2dSn_Y}, respectively. The gray solid line again corresponds to an analytical CW solution. As for the case $S=0$, the tilted solutions drift not only in the longitudinal, but also in the transversal $x$-direction. Although three-peak tilted LSs were unstable for $S=0$ for the given value of $B$, it becomes stabilized in the presence of $S$. However, as in the case $S=0$, no stable dark LSs were found for this parameters set. To study the stability range of the found LSs, we performed a numerical scan in the diffraction $B$ for fixed values of the injection $Y$ and aberration $S$, see Fig.~\ref{fig:2dSn_Y} (right). By varying $B$, one can see that all the LS solutions presented in  Fig.~\ref{fig:2dSn_Y} (left), exist and are stable in a wide range of the diffraction values. In particular, increasing $B$ and following the branch of the straight two-peak LSs (blue), one notices that it becomes unstable with respect to Andronov-Hopf (AH) bifurcation, and an oscillating, straight two-peak LS is formed (brawn). This solution becomes unstable by further increasing $B$ and one jumps to a tilted two-peak LSs (black), that is stable and co-exists with both the stationary and oscillating two-peak LSs.

Remarkably, LSs remain stable even for $B < 0$. In this regime, paraxial diffraction alone cannot provide transverse confinement, so the transverse localization is entirely due to the fourth-order diffraction. Since the transverse operator in Eq.~\eqref{eq:LLE_norm} contributes a phase rate $-i\,(B\,k^2 - S\,k^4)$, the term $S\,k^4$ for $S<0$ goes up like $B\,k^2$ when $B>0$. This demonstrates that the aberration is not a small correction but the confinement mechanism itself. Indeed, the results of the numerical scan for decreasing $B$ indicates that both two-peak straight and four-peak tilted LSs remain stable in a range of negative $B$ values. Further, the three-peak tilted LSs becomes unstable close to $B=0$, leading to a formation of an 3-peak oscillating LSs (orange). The latter stabilizes in a AH bifurcation and a stable straight three-peak LS is created (green).
\begin{figure}[h!]
 \centering
 \includegraphics[width=0.33\columnwidth]{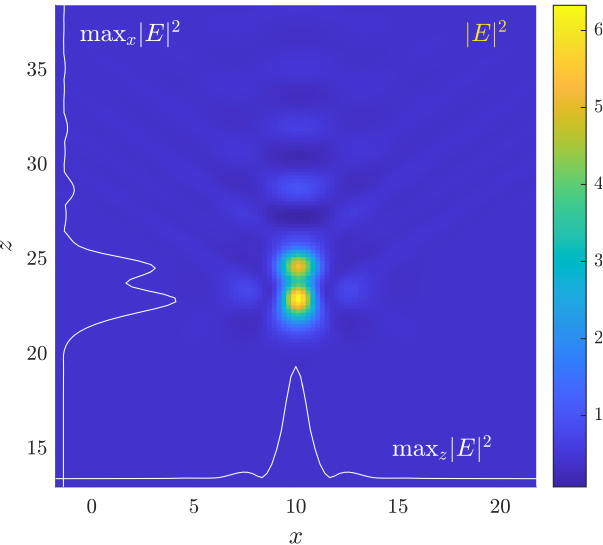}~\includegraphics[width=0.33\columnwidth]{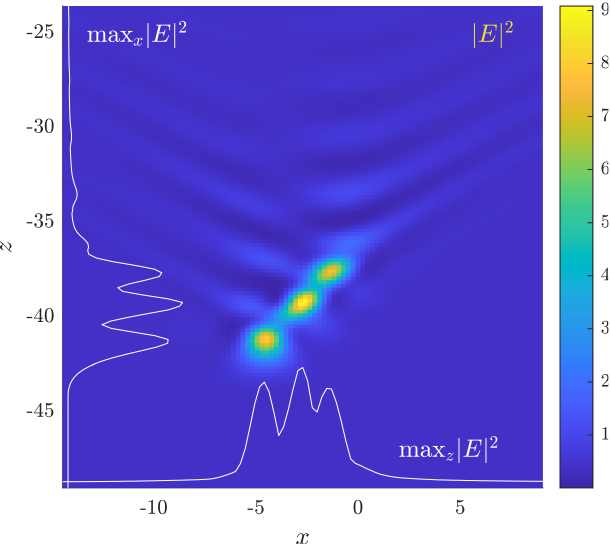}~\includegraphics[width=0.33\columnwidth]{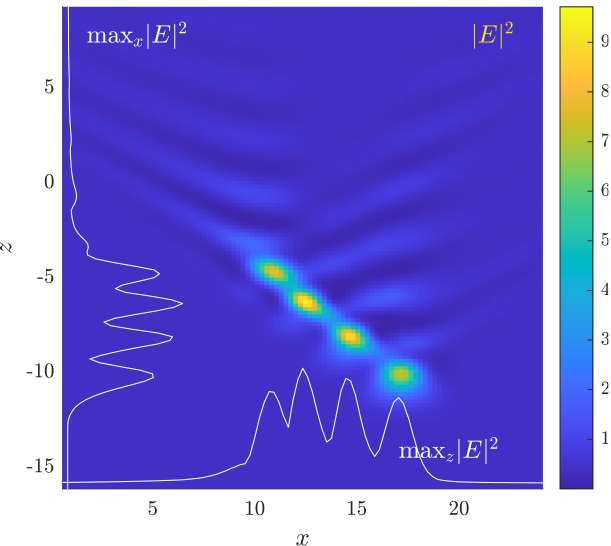}\\
  \includegraphics[width=0.33\columnwidth]{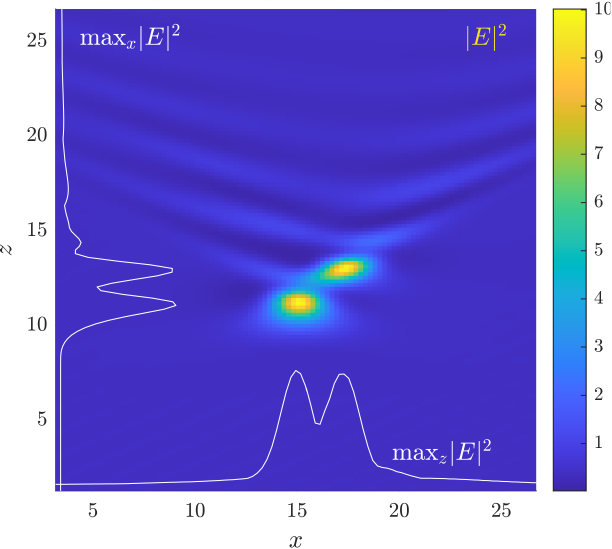}~\includegraphics[width=0.33\columnwidth]{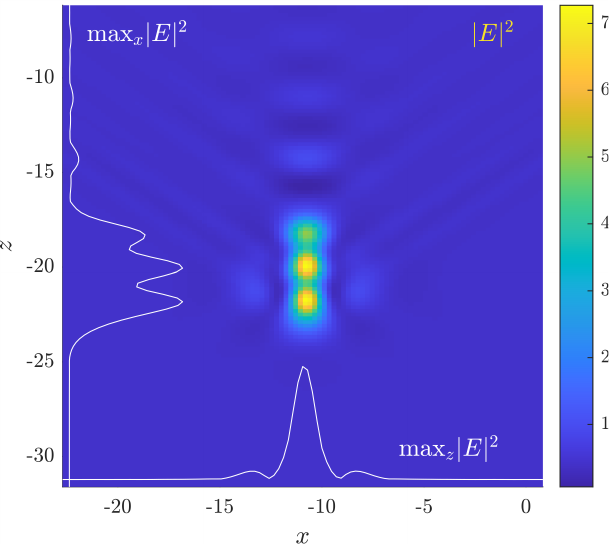}~\includegraphics[width=0.33\columnwidth]{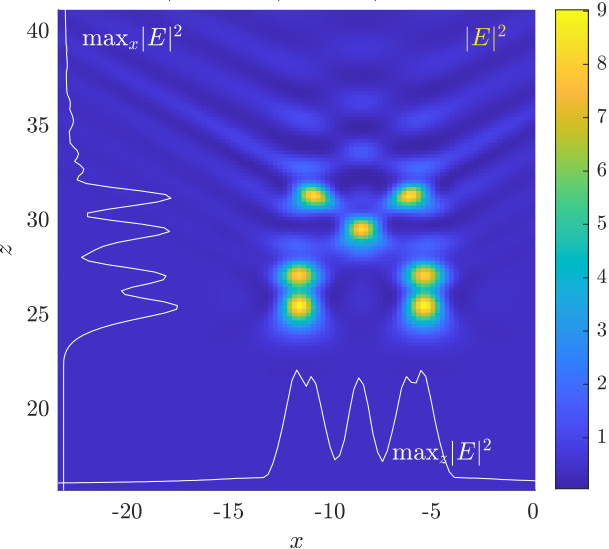}
 \caption{Exemplary two-dimensional LSs obtained in the numerical integration of Eq.~\eqref{eq:LLE_norm} in the normal dispersion regime at $Y=2.38$ and $S=-0.5$ for different values of  $B=\{-1.186,\,0.168,\,0.6,\,-1.02,\,4.405,\,0.3\}$ (left to right, top to bottom, respectively). A zoom into the spatial intensity distribution $|E(x,z)|^2$ is shown together with projections in $x$ and $z$ onto the respective axis in white, indicating the intensity maximum over the corresponding axis. Other parameters as in Fig.~\ref{fig:2dS0_YB}.
 }\label{fig:2dS_profiles}
 \end{figure}
Figure~\ref{fig:2dS_profiles} shows typical exemplary intensity profiles $|E(x,z)|^2$ of all non-oscillatory LS solutions found in Fig.~\ref{fig:2dSn_Y} (right), whereas a Supplementary Video I presents a visualization of an oscillating three-peak LS, obtained at $B=-0.36$. One can see that the three-peaked LS uniformly moves in the longitudinal direction while undergoing a pronounced lateral undulation, or wiggling, in the $x$-direction. This transverse wiggling is characterized by a periodic shift of the entire multi-peak envelope, causing it to weave back and forth as it translates.

Although no stable straight or tilded two-dimensional LSs consisting of more than four peaks were found during our numerical scans, we found complex bound states of several LSs that can be stable in a large parameter range of $B$, see the pink branch in Fig. Fig.~\ref{fig:2dSn_Y} (right). This composed state consist of seven LSs, where a pair of straight two-peak LSs is bounded with a pair of tilted two-peak solution, see Fig.~\ref{fig:2dS_profiles}~(bottom, right) for an exemplary profile. The bound state uniformly moves in the $z$-direction with no transversal drift, see Supplementary Video II for a visualization.

\begin{figure}[h!]
 \includegraphics[width=0.99\columnwidth]{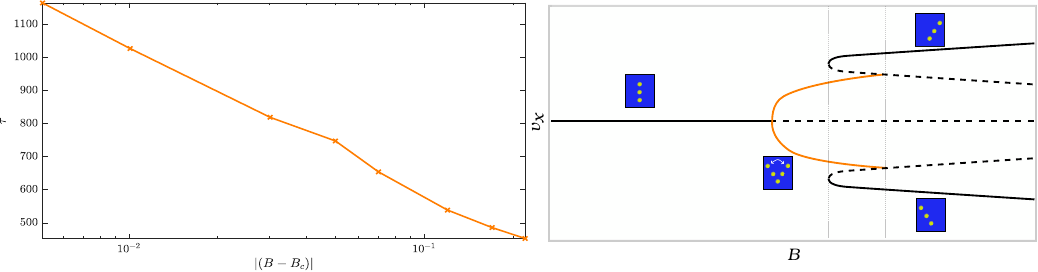}\caption{(Left) The period $\tau$ of the oscillation over the absolute
distance to the critical point $B_c$. Parameters as in Fig.~\ref{fig:2dS0_YB}. (Right) Schematic of a possible bifurcation scenario of the 3-peak LS.  We present the bifurcation diagram for the transverse ve-
locity $v_x$ as a function of $B$, showing the transition from stable straight LS to an oscillating LS in a supercritical AH bifurcation and to one of the tilted LS in a homoclinic bifurcation. Grey dotted lines indicate the transitions in between the branches as part of the observed hysteresis.}\label{fig:2dB_hombif}
\end{figure}

In the next step we turn our attention to the bifurcation scenario for a three-peak straight LS (cf. green line in  Fig.~\ref{fig:2dSn_Y} (right)), observed at small $B$ values. The straight three-peak LS uniformly drifts in the $z$-direction without the transversal drift. It undergoes an AH bifurcation at $B=B_H \simeq -0.9$ and, close to the bifurcation point, the amplitude of the oscillations increases as $\mathcal{O}(\sqrt{B-B_H})$, indicating a supercritical bifurcation. However, the branch of oscillating LSs (cf. orange line in Fig.~\ref{fig:2dSn_Y} (right)) suddenly disappears at $B=B_c\simeq -0.22$ and the solution falls on to one of the branches of the right- or left-tilted three peak LS (cf. red line in Fig.~\ref{fig:2dSn_Y} (right), here right- and left-tilted solutions are non-distinguishable for the chosen norm). The down-sweep in $B$ indicates a hysteresis behaviour for the tilded LS, where the solution becomes unstable only at $B=B_{h} \simeq -0.415$. We can also track the period $\tau$ of the oscillating LS close to $B_c$, see Fig.~\ref{fig:2dB_hombif}~(left).
As the horizontal axis is scaled logarithmically, the data displays a linear decline, confirming the scaling law typical for a homoclinic bifurcation. All these indicators suggest a bifurcation scenario that is laid out schematically in Fig.~\ref{fig:2dB_hombif}~(right), where we schematically depict the bifurcation diagram for the transverse ve-
locity $v_x$ as a function of $B$, showing the transition from stable straight LS to an oscillating LS in a supercritical AH bifurcation and to one of the tilted LS in a homoclinic bifurcation, there the limit cycle collides with the corresponding saddles. However, we cannot prove the existence of these unstable branches as we only employ time simulations. A similar bifurcation scenario can be suggested for the transition between an straight and tilted two-peak LSs observed at large $B$ values (cf. Fig.~\ref{fig:2dSn_Y}~(right)).
\begin{figure}[h!]
 \includegraphics[width=0.99\columnwidth]{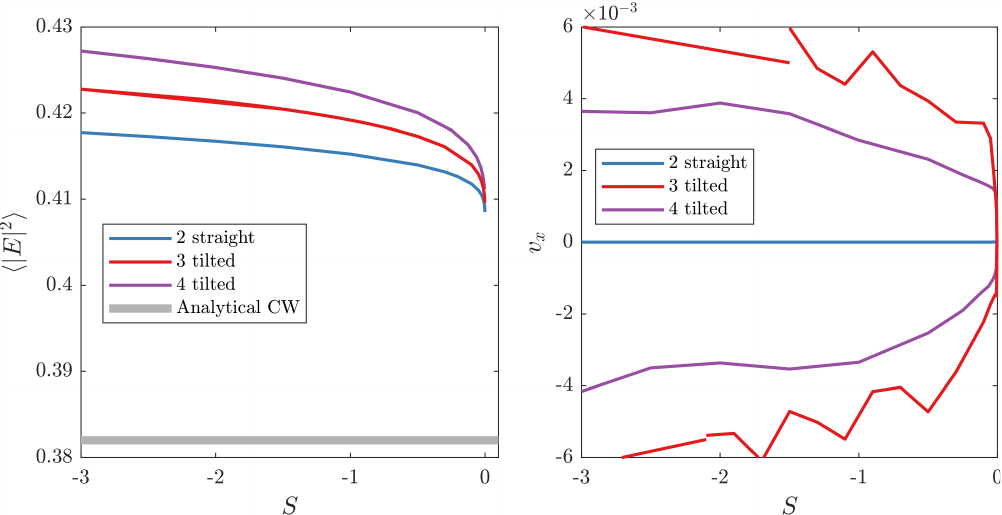}
 \caption{ Bifurcation diagram obtained by a numerical integration of a 2D Eq.~\eqref{eq:LLE_norm} for $Y=2.32$ and $B=0.01$. (Left) The average field intensity and (right) transverse velocity as a function of the aberration $S$ is presented. Blue, red and violet lines correspond to the straight two-peak LSs, tilted three- and four-peak LSs, respectively. The gray line corresponds to the analytically calculated CW branch, where the stability information is omitted. Other parameters as in Fig.~\ref{fig:2dS0_YB}.}\label{fig:2dS}
\end{figure}

Finally, we examine the effect of the aberration $S$ on the shape and stability of the LSs. To this aim we fixed the injection $Y$ and the diffraction $B$ to the values, where the three LSs types (straight two-peak LS as well as tilted three- and four-peak LSs are multistable, cf. Fig.~\ref{fig:2dSn_Y}) and performed a numerical scan in $S$, see Fig.~\ref{fig:2dS}, where the evolution of the average intensity (left) and the transverse velocity (right) is presented. As $|S|\rightarrow 0$, the transverse velocity vanishes and the average intensity approaches the CW background, the excess intensity scaling as $\sqrt[4]{|S|}$ for all LS types. The structures thus shrink transversally and disappear as the aberration vanishes, indicating that the transverse scale is set by the fourth-order diffraction.

\section{Dynamics of localized states in three dimensions}

In this section, we extend our numerical analysis into three dimensions, introducing additional transversal dimension, i.e., in Eq.~\eqref{eq:LLE_norm} $E=E(x,\,y,\,z,\,t)$ and $\Delta_{\perp}=\partial_x^2+\partial_y^2$. We follow the same scanning protocol as in the 2D case, discussed above. We solve Eq.~\eqref{eq:LLE_norm} numerically on the rectangular domain $(x,\,y,\,z)\in \Omega=[-30,\,30]\times[-30,\,30] \times[-50,\,50]$ with periodic boundary conditions employing a split-step Fourier method subjected to noise of a small amplitude. As in the 2D case, up- and down- sweeps are performed during all the parameter scans.

\begin{figure}[h!]
 \centering
 \includegraphics[width=0.44\columnwidth]{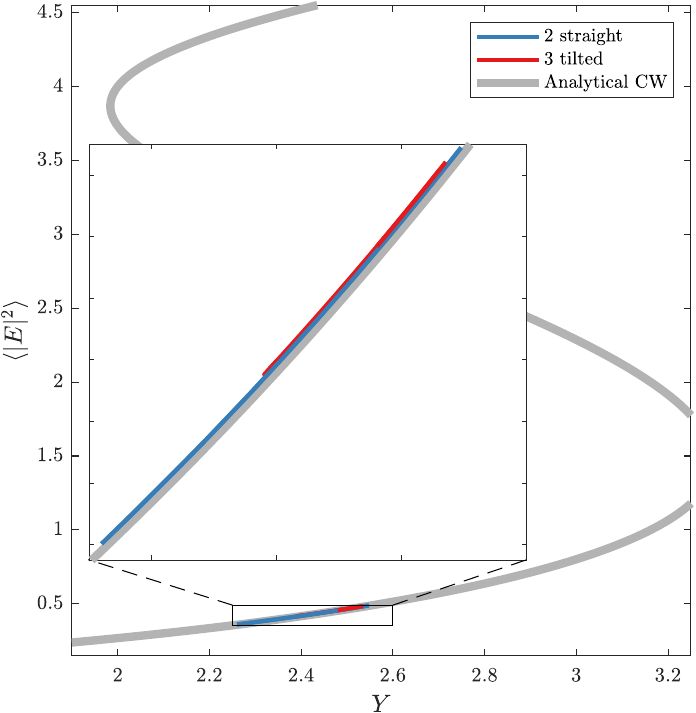}~\includegraphics[width=0.45\columnwidth]{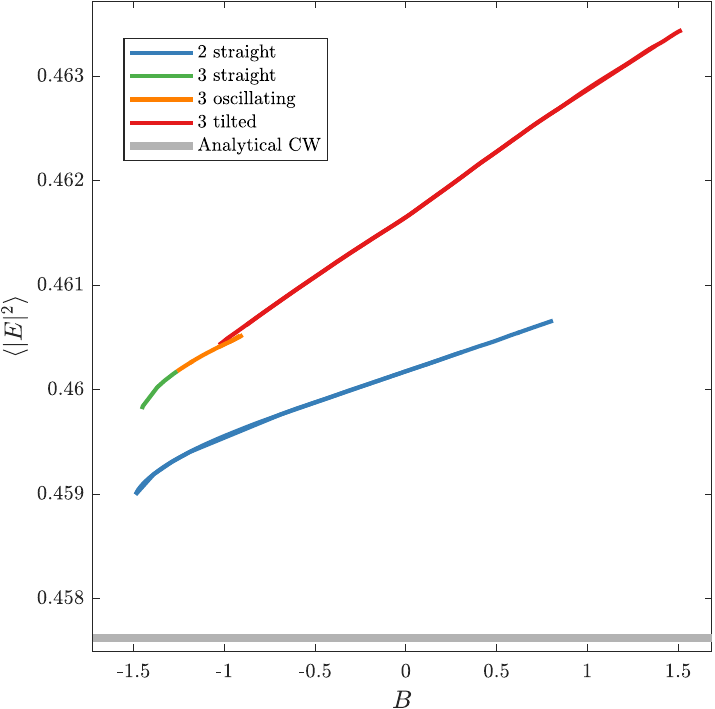}
 \caption{Bifurcation diagram obtained by numerical integration of a 3D Eq.~\eqref{eq:LLE_norm} for $S=-0.5$. The average field intensity as a function of (left) injection $Y$ and (right) diffraction $B$ calculated at $B=0.01$ (left) and $Y=2.49$ (right) is presented. Blue and green lines correspond to an straight two- and three- peak LSs, red and violet lines represent tilted three- and four-peak LSs, respectively.Three- peak oscillating LSs are depicted in orange. The gray line corresponds to the analytically calculated CW branch, where the stability information is omitted. Other parameters as in Fig.~\ref{fig:2dS0_YB}.
 }\label{fig:3dSn_Y}
\end{figure}

To study the impact of an additional transversal dimension on the stability of the three-dimensional LSs, the light bullets (LBs),  we start our analysis with a numerical scan in the injection $Y$, keeping the same values of $B=0.01$ and $S=-0.5$ as in 2D (cf. Fig.~\ref{fig:2dSn_Y}). The results are presented in Fig.~\ref{fig:3dSn_Y} (left). We identified three-dimensional counterparts of the LSs found in the 2D scan: The two-peak straight LB (blue line) and a tilted three-peak LB (red line). Despite extensive searches, we did not find a four-peak LB. It may lose stability with the extra dimension or be inherently difficult to locate.
We note that the parameter scan results are similar to the 2D results. We observe a potential snaking structure, suggesting saddle-node bifurcations at both ends of the two branches. Surprisingly, the stability range of the two-peak straight LB is larger in 3D, as the LB is stable at smaller $Y$ values as compared to the 2D case.
\begin{figure}[h!]
  \includegraphics[width=0.85\columnwidth]{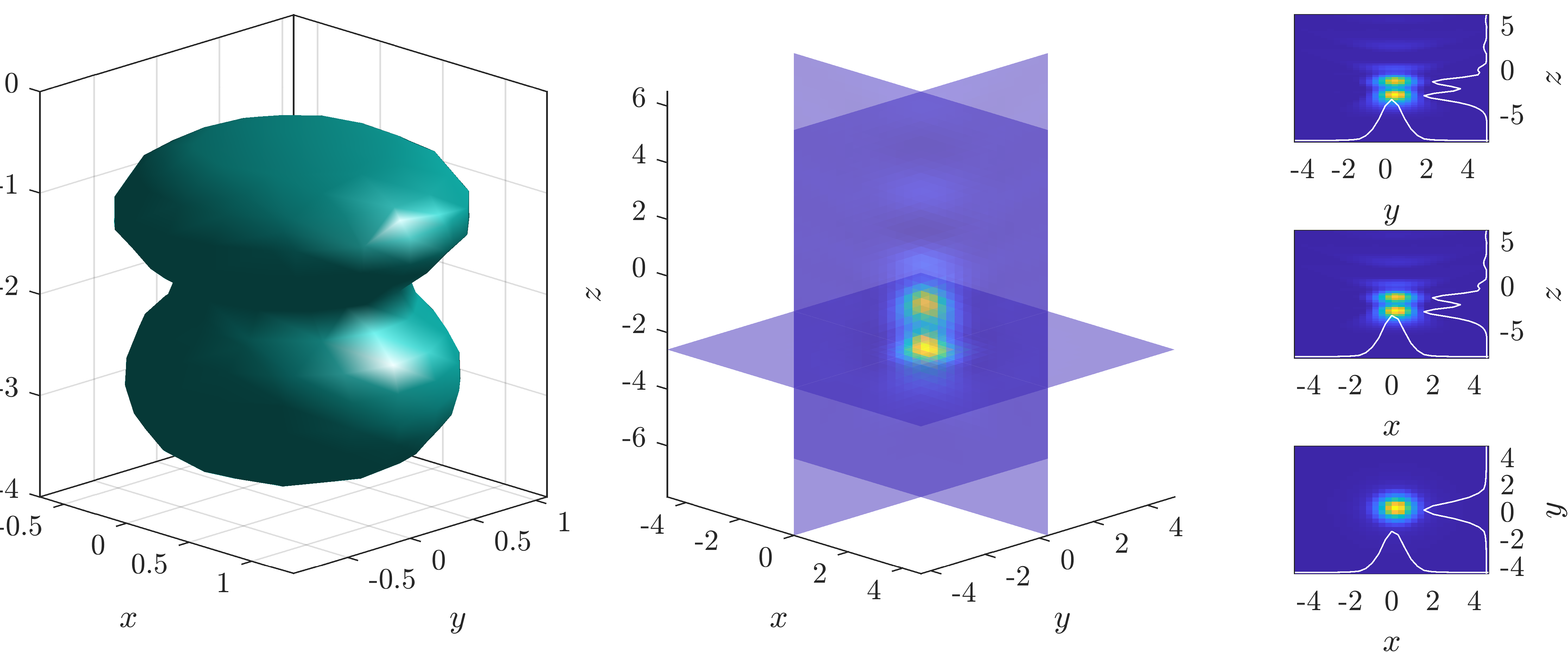}\\
  \includegraphics[width=0.85\columnwidth]{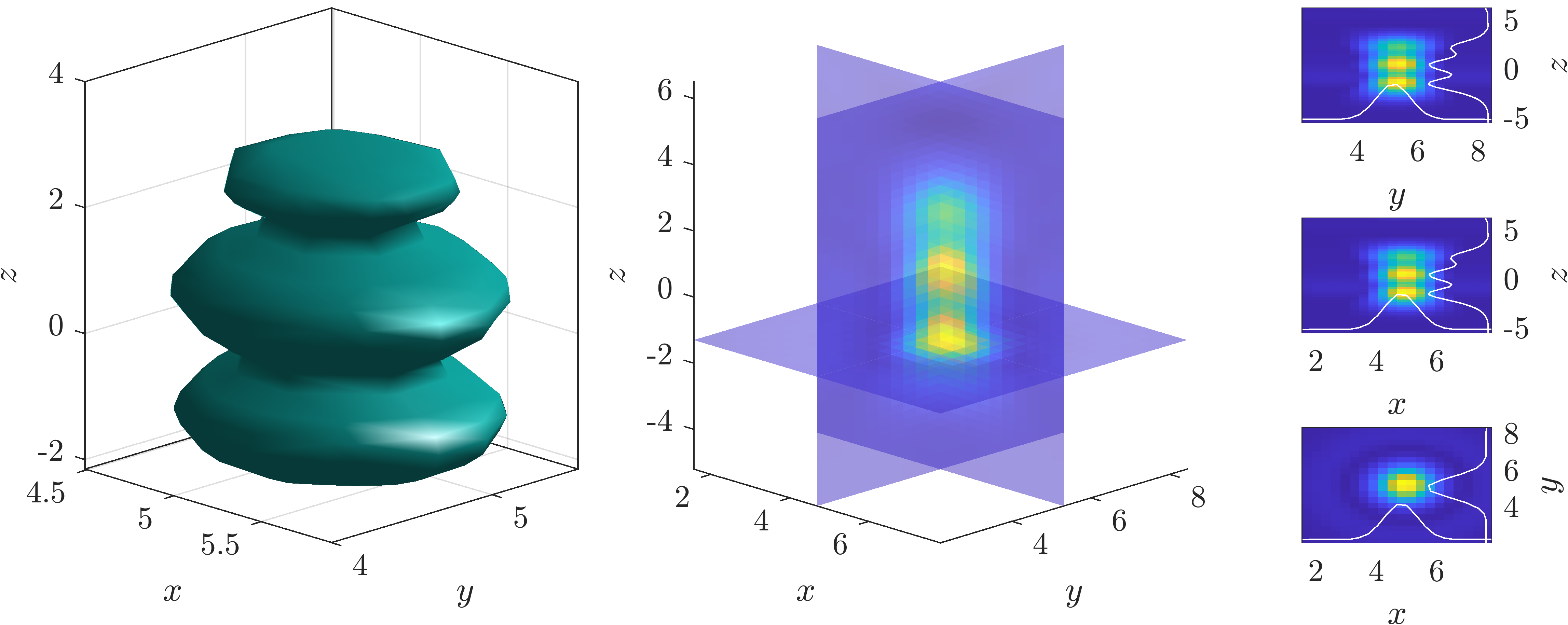}\\
  \includegraphics[width=0.85\columnwidth]{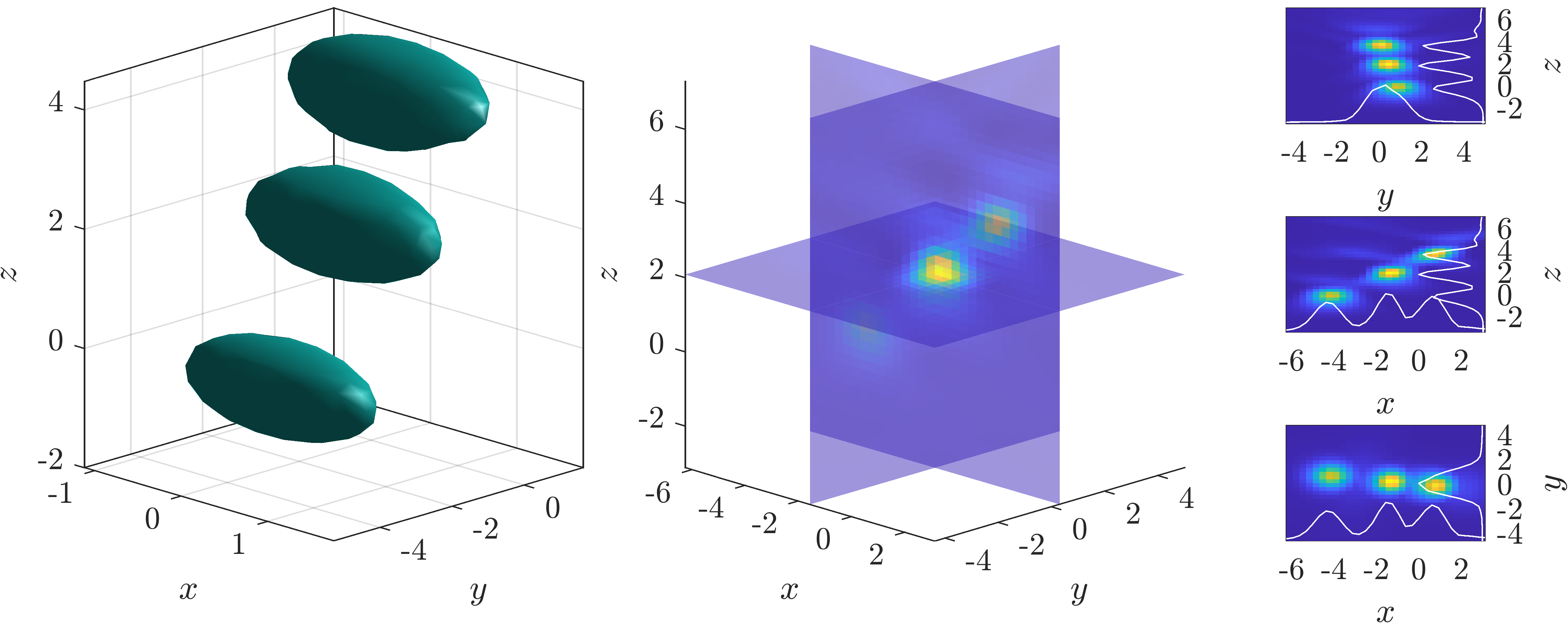}
  \caption{Exemplary three-dimensional profiles of two-peak (top) and three-peak (middle) straight LBs as well as of a tilted three-peak LB (bottom) obtained in the numerical integration of Eq.~\eqref{eq:LLE_norm} in the normal dispersion regime for $Y=2.49$ and $S=-0.5$. (Left) An isosurface plot of the intensity $|E(x,y,z)|^2$, with a cutoff value 1 to 10 $\%$ above the intensity of the background state. (Center) Cross-sections of $|E(x,y,z)|^2$ along each axis, with semi-transparent color maps centered on the coordinate of maximum intensity. (Right) Intensity maxima across each dimension in 2D, with a white lines indicating the maximum across the remaining axis. Parameter are $B=\{0.1,\,-1.455,\,0.615\}$ (top to bottom, respectively). Other parameters as in Fig.~\ref{fig:2dS0_YB}.}\label{fig:3dS_profiles}
\end{figure}
Further, we fixed the value of $Y$ where both found LB are stable and conducted a numerical scan in the diffraction $B$, see Fig.~\ref{fig:3dSn_Y}~(right). The scan revealed three-dimensional LB counterparts to the two-dimensional straight two- and three- peak LSs (blue and green lines), tilted three-peak LS (red) as well as oscillating three-peak LS (orange) for negative diffraction values. The branches are organized in a similar way and seem to be connected in the similar way. However, we were not able to find the oscillating and tilted two-peak LB. Another point to emphasize is that the range of the stability in $B$ of all found LBs is smaller than in 2D and is limited to small values of $|B|$. We conclude that, in 3D, LBs exist only where the aberration dominates over paraxial diffraction, i.e. close to the self-imaging condition, which underlines that the stabilization relies on the fourth-order term.
\begin{figure}[h!]
\centering
 \includegraphics[width=0.83\columnwidth]{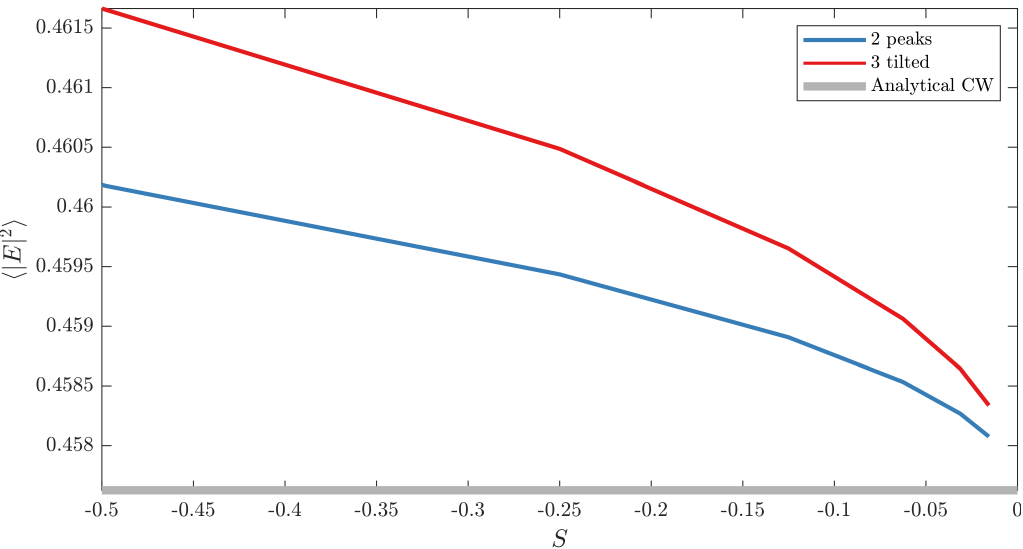}
 \caption{Bifurcation diagram obtained by a numerical integration of a 3D Eq.~\eqref{eq:LLE_norm} for $Y=2.49$ and $B=0.01$.  The average field intensity as a function of the aberration $S$ is presented. Blue and red  lines correspond to the straight two-peak LB and tilted three-peak LB, respectively. The gray line corresponds to the analytically calculated CW branch, where the stability information is omitted Other parameters as in Fig.~\ref{fig:2dS0_YB}}\label{fig:3dS}
\end{figure}
The exemplary profiles of two- and three-peak straight LBs as well as of a tilted three-peak LB are presented in Fig.~\ref{fig:3dS_profiles} (top to bottom, respectively). Here, three-dimensional stationary intensity distributions $|E(x,y,z)|^2$ are depicted with three representations: left panels shows an isosurface plot of the intensity, with a cutoff value 1 to 10 $\%$ above the intensity of the background state. The center panels displays cross-sections of $|E(x,y,z)|^2$ along each axis, with semi-transparent color maps centered on the coordinate of maximum intensity. On the right, three plots show intensity maxima across each dimension in 2D, with a white lines indicating the maximum across the remaining axis. Further, the Supplementary Video III illustrates the dynamics of the oscillating three-peak LB (cf. orange branch in Fig.~\ref{fig:3dSn_Y}~(right)). We note that as in 2D case, straight two- and three-peaks LBs uniformly propagate in the longitudinal direction, while the both transversal speeds $v_x=v_y=0$. In contrast, the tilted LB propagates with $v_x,\,v_y\neq0$. While in two-dimensional case, only two tilted configurations were possible, infinite set of tilted states exists, since the the tilt angle toward the $z$-axis remains fixed, the angle within the $(x, y)$ plane is arbitrary.
Finally, we test the range of the stability of obtained LBs by changing the aberration parameter $S$. The results are presented in Fig.~\ref{fig:3dS}, where we followed the branches of two-peak LB and tilted three-peak LB. One can see that the LBs are stable in a large range of $S$. The
intensity decreases sharply as $S\rightarrow 0$, compressing the states in the transversal directions. This transverse compression for vanishing aberration is precisely the tendency towards collapse that the fourth-order diffraction counteracts. We note that we were not able to find any stable LB solution for $S=0$ for a given parameter set.

\section{Conclusion}
In this work, we studied the formation and stability of multidimensional localized states in an injected Kerr–Gires–Tournois interferometer in the normal dispersion regime. Close to the self-imaging condition, fourth-order diffraction induced by spherical aberrations becomes relevant. Our analysis was based on a generalized spatio-temporal Lugiato–Lefever model including third-order dispersion, paraxial diffraction, and spherical aberration.

In two dimensions and in the absence of aberrations, only tilted two- and three-peak localized states were found to be stable. These states drift both longitudinally and transversely. Neither dark states nor straight bright states were observed. Introducing a small negative aberration drastically changes this picture. A multistable set of straight two-peak, tilted three- and four-peak, and oscillating states, as well as complex bound states, becomes stable over wide ranges of injection and diffraction, even for negative paraxial diffraction. Following these branches revealed a supercritical Andronov–Hopf bifurcation leading to transversely wiggling states. The oscillation period scales logarithmically near the critical point, indicating a homoclinic bifurcation toward the tilted states, accompanied by hysteresis. In three dimensions, we found stable straight two- and three-peak light bullets, tilted three-peak light bullets, and oscillating three-peak light bullets. Since the tilt angle in the transverse plane is arbitrary, the tilted bullets form an infinite family. Their stability range in diffraction is narrower than in 2D, and no stable light bullets were found for vanishing aberration. We conclude that spherical aberration is therefore an essential stabilizing ingredient.

An important limitation of the present study is that it relies exclusively on direct numerical integration. This approach gives access only to stable solutions and the information about unstable states as well as  about how the different branches are connected, remains unclear. The saddle-node bifurcations suggested at the branch ends, the collapsed snaking structure, and the unstable saddle branches involved in the proposed homoclinic scenario can be inferred but not demonstrated.

A natural way to overcome this difficulty is numerical path continuation. However, applying existing frameworks such as pde2path~\cite{uecker2014} is far from straightforward here. Indeed, pde2path is based on the finite element method, so the third-order dispersion term and the fourth-order transverse operator must be split into many auxiliary fields. With the spatial resolution required to resolve the oscillatory tails and multi-peak structures, especially in three dimensions, the resulting system becomes prohibitively large. Fourier-based continuation methods would be much better suited, since all linear operators of the model are diagonal in Fourier space and no auxiliary fields are required. However, the current Fourier-based implementation in pde2path is restricted to even-order operators. It therefore cannot directly handle the odd third-order dispersion term needed to describe drifting and tilted states. Extending Fourier-based continuation frameworks to general, including odd, differential operators would allow a complete bifurcation analysis of the unstable states and of the organization of multidimensional light bullets. This is left for future work.

\section*{Acknowledgments}
J.J. acknowledges the financial support of the Ministerio de Economía y Competitividad (PID2021-128910NB-100 AEI/FEDER UE). J.J. and S.V.G. acknowledge the financial support of the project KOGIT, Agence Nationale de la Recherche (ANR-22-CE92-0009), Deutsche Forschungsgemeinschaft (DFG), Germany, via Grant Nos. 505936983 and 524947050.

\section*{Declaration of competing interest}
The authors declare that they have no known competing financial interests or personal relationships that could have appeared to influence the work reported in this paper.

\begin{thebibliography}{58}
\expandafter\ifx\csname natexlab\endcsname\relax\def\natexlab#1{#1}\fi
\providecommand{\url}[1]{\texttt{#1}}
\providecommand{\href}[2]{#2}
\providecommand{\path}[1]{#1}
\providecommand{\DOIprefix}{doi:}
\providecommand{\ArXivprefix}{arXiv:}
\providecommand{\URLprefix}{URL: }
\providecommand{\Pubmedprefix}{pmid:}
\providecommand{\doi}[1]{\href{http://dx.doi.org/#1}{\path{#1}}}
\providecommand{\Pubmed}[1]{\href{pmid:#1}{\path{#1}}}
\providecommand{\bibinfo}[2]{#2}
\ifx\xfnm\relax \def\xfnm[#1]{\unskip,\space#1}\fi
\bibitem[{Sch\"apers et~al.(2000)Sch\"apers, Feldmann, Ackemann, and
  Lange}]{SFA-PRL-00}
\bibinfo{author}{B.~Sch\"apers}, \bibinfo{author}{M.~Feldmann},
  \bibinfo{author}{T.~Ackemann}, \bibinfo{author}{W.~Lange},
\newblock \bibinfo{title}{Interaction of localized structures in an optical
  pattern-forming system},
\newblock \bibinfo{journal}{Phys. Rev. Lett.} \bibinfo{volume}{85}
  (\bibinfo{year}{2000}) \bibinfo{pages}{748--751}.
  \DOIprefix\doi{10.1103/PhysRevLett.85.748}.
\bibitem[{Barland et~al.(2002)Barland, Tredicce, Brambilla, Lugiato, Balle,
  Giudici, Maggipinto, Spinelli, Tissoni, Kn{\"o}dl, Miller, and
  J{\"a}ger}]{BTB-NAT-02}
\bibinfo{author}{S.~Barland}, \bibinfo{author}{J.~R. Tredicce},
  \bibinfo{author}{M.~Brambilla}, \bibinfo{author}{L.~A. Lugiato},
  \bibinfo{author}{S.~Balle}, \bibinfo{author}{M.~Giudici},
  \bibinfo{author}{T.~Maggipinto}, \bibinfo{author}{L.~Spinelli},
  \bibinfo{author}{G.~Tissoni}, \bibinfo{author}{T.~Kn{\"o}dl},
  \bibinfo{author}{M.~Miller}, \bibinfo{author}{R.~J{\"a}ger},
\newblock \bibinfo{title}{Cavity solitons as pixels in semiconductor
  microcavities},
\newblock \bibinfo{journal}{Nature} \bibinfo{volume}{419}
  (\bibinfo{year}{2002}) \bibinfo{pages}{699--702}.
  \DOIprefix\doi{10.1038/nature01049}.
\bibitem[{Pesch et~al.(2007)Pesch, Lange, Gomila, Ackemann, Firth, and
  Oppo}]{PLG-PRL-08}
\bibinfo{author}{M.~Pesch}, \bibinfo{author}{W.~Lange},
  \bibinfo{author}{D.~Gomila}, \bibinfo{author}{T.~Ackemann},
  \bibinfo{author}{W.~J. Firth}, \bibinfo{author}{G.-L. Oppo},
\newblock \bibinfo{title}{Two-dimensional front dynamics and spatial solitons
  in a nonlinear optical system},
\newblock \bibinfo{journal}{Phys. Rev. Lett.} \bibinfo{volume}{99}
  (\bibinfo{year}{2007}) \bibinfo{pages}{153902}.
  \DOIprefix\doi{10.1103/PhysRevLett.99.153902}.
\bibitem[{Ackemann et~al.(2009)Ackemann, Firth, and Oppo}]{AFO-AAM-09}
\bibinfo{author}{T.~Ackemann}, \bibinfo{author}{W.~J. Firth},
  \bibinfo{author}{G.~Oppo},
\newblock \bibinfo{title}{Chapter 6 fundamentals and applications of spatial
  dissipative solitons in photonic devices},
\newblock in: \bibinfo{editor}{P.~R.~B. E.~Arimondo}, \bibinfo{editor}{C.~C.
  Lin} (Eds.), \bibinfo{booktitle}{Advances in Atomic Molecular and Optical
  Physics}, volume~\bibinfo{volume}{57} of \textit{\bibinfo{series}{Advances In
  Atomic, Molecular, and Optical Physics}}, \bibinfo{publisher}{Academic
  Press}, \bibinfo{year}{2009}, pp. \bibinfo{pages}{323 -- 421}.
  \DOIprefix\doi{http://dx.doi.org/10.1016/S1049-250X(09)57006-1}.
\bibitem[{Gustave et~al.(2017)Gustave, Radwell, McIntyre, Toomey, Kane,
  Barland, Firth, Oppo, and Ackemann}]{GRMc-PRL-17}
\bibinfo{author}{F.~Gustave}, \bibinfo{author}{N.~Radwell},
  \bibinfo{author}{C.~McIntyre}, \bibinfo{author}{J.~P. Toomey},
  \bibinfo{author}{D.~M. Kane}, \bibinfo{author}{S.~Barland},
  \bibinfo{author}{W.~J. Firth}, \bibinfo{author}{G.-L. Oppo},
  \bibinfo{author}{T.~Ackemann},
\newblock \bibinfo{title}{Observation of mode-locked spatial laser solitons},
\newblock \bibinfo{journal}{Phys. Rev. Lett.} \bibinfo{volume}{118}
  (\bibinfo{year}{2017}) \bibinfo{pages}{044102}.
  \DOIprefix\doi{10.1103/PhysRevLett.118.044102}.
\bibitem[{Ivars et~al.(2023)Ivars, Kartashov, de~C{\'o}rdoba, Conejero, Torner,
  and Mili{\'a}n}]{IKC-NatPhot-23}
\bibinfo{author}{S.~B. Ivars}, \bibinfo{author}{Y.~V. Kartashov},
  \bibinfo{author}{P.~F. de~C{\'o}rdoba}, \bibinfo{author}{J.~A. Conejero},
  \bibinfo{author}{L.~Torner}, \bibinfo{author}{C.~Mili{\'a}n},
\newblock \bibinfo{title}{Photonic snake states in two-dimensional frequency
  combs},
\newblock \bibinfo{journal}{Nature Photonics} \bibinfo{volume}{17}
  (\bibinfo{year}{2023}) \bibinfo{pages}{767--774}.
  \DOIprefix\doi{10.1038/s41566-023-01220-1}.
\bibitem[{Kartashov et~al.(2019)Kartashov, Astrakharchik, Malomed, and
  Torner}]{KAM-NatRevPhys-19}
\bibinfo{author}{Y.~V. Kartashov}, \bibinfo{author}{G.~E. Astrakharchik},
  \bibinfo{author}{B.~A. Malomed}, \bibinfo{author}{L.~Torner},
\newblock \bibinfo{title}{Frontiers in multidimensional self-trapping of
  nonlinear fields and matter},
\newblock \bibinfo{journal}{Nature Reviews Physics} \bibinfo{volume}{1}
  (\bibinfo{year}{2019}) \bibinfo{pages}{185--197}.
  \DOIprefix\doi{10.1038/s42254-019-0025-7}.
\bibitem[{Silberberg(1990)}]{S-OL-90}
\bibinfo{author}{Y.~Silberberg},
\newblock \bibinfo{title}{Collapse of optical pulses},
\newblock \bibinfo{journal}{Opt. Lett.} \bibinfo{volume}{15}
  (\bibinfo{year}{1990}) \bibinfo{pages}{1282--1284}.
\bibitem[{{Zakharov}(1972)}]{Z-SJETP-72}
\bibinfo{author}{V.~E. {Zakharov}},
\newblock \bibinfo{title}{{Collapse of Langmuir Waves}},
\newblock \bibinfo{journal}{Soviet Journal of Experimental and Theoretical
  Physics} \bibinfo{volume}{35} (\bibinfo{year}{1972}) \bibinfo{pages}{908}.
\bibitem[{Rasmussen and Rypdal(1986)}]{RR-PhScr-86}
\bibinfo{author}{J.~J. Rasmussen}, \bibinfo{author}{K.~Rypdal},
\newblock \bibinfo{title}{Blow-up in nonlinear schroedinger equations-i a
  general review},
\newblock \bibinfo{journal}{Physica Scripta} \bibinfo{volume}{33}
  (\bibinfo{year}{1986}) \bibinfo{pages}{481}.
  \DOIprefix\doi{10.1088/0031-8949/33/6/001}.
\bibitem[{Yu et~al.(1995)Yu, Chien, Lai, and Wang}]{YCL-OptCom-95}
\bibinfo{author}{S.-S. Yu}, \bibinfo{author}{C.-H. Chien},
  \bibinfo{author}{Y.~Lai}, \bibinfo{author}{J.~Wang},
\newblock \bibinfo{title}{Spatio-temporal solitary pulses in graded-index
  materials with kerr nonlinearity},
\newblock \bibinfo{journal}{Optics Communications} \bibinfo{volume}{119}
  (\bibinfo{year}{1995}) \bibinfo{pages}{167--170}.
  \DOIprefix\doi{https://doi.org/10.1016/0030-4018(95)00377-K}.
\bibitem[{Tlidi and Mandel(1999)}]{TM-PRL-99}
\bibinfo{author}{M.~Tlidi}, \bibinfo{author}{P.~Mandel},
\newblock \bibinfo{title}{Three-dimensional optical crystals and localized
  structures in cavity second harmonic generation},
\newblock \bibinfo{journal}{Phys. Rev. Lett.} \bibinfo{volume}{83}
  (\bibinfo{year}{1999}) \bibinfo{pages}{4995--4998}.
  \DOIprefix\doi{10.1103/PhysRevLett.83.4995}.
\bibitem[{Veretenov et~al.(2000)Veretenov, Vladimirov, Kaliteevskii, Rozanov,
  Fedorov, and Shatsev}]{VVK-OS-00}
\bibinfo{author}{N.~Veretenov}, \bibinfo{author}{A.~Vladimirov},
  \bibinfo{author}{N.~Kaliteevskii}, \bibinfo{author}{N.~Rozanov},
  \bibinfo{author}{S.~Fedorov}, \bibinfo{author}{A.~Shatsev},
\newblock \bibinfo{title}{Conditions for the existence of laser bullets},
\newblock \bibinfo{journal}{Optics and Spectroscopy} \bibinfo{volume}{89}
  (\bibinfo{year}{2000}) \bibinfo{pages}{380--383}.
  \DOIprefix\doi{10.1134/1.1310704}.
\bibitem[{Bang et~al.(2002)Bang, Krolikowski, Wyller, and
  Rasmussen}]{BKW-PRE-02}
\bibinfo{author}{O.~Bang}, \bibinfo{author}{W.~Krolikowski},
  \bibinfo{author}{J.~Wyller}, \bibinfo{author}{J.~J. Rasmussen},
\newblock \bibinfo{title}{Collapse arrest and soliton stabilization in nonlocal
  nonlinear media},
\newblock \bibinfo{journal}{Phys. Rev. E} \bibinfo{volume}{66}
  (\bibinfo{year}{2002}) \bibinfo{pages}{046619}.
  \DOIprefix\doi{10.1103/PhysRevE.66.046619}.
\bibitem[{Brambilla et~al.(2004)Brambilla, Maggipinto, Patera, and
  Columbo}]{BMP-PRL-04}
\bibinfo{author}{M.~Brambilla}, \bibinfo{author}{T.~Maggipinto},
  \bibinfo{author}{G.~Patera}, \bibinfo{author}{L.~Columbo},
\newblock \bibinfo{title}{Cavity light bullets: Three-dimensional localized
  structures in a nonlinear optical resonator},
\newblock \bibinfo{journal}{Phys. Rev. Lett.} \bibinfo{volume}{93}
  (\bibinfo{year}{2004}) \bibinfo{pages}{203901}.
\bibitem[{Columbo et~al.(2006)Columbo, Perrini, Maggipinto, and
  Brambilla}]{CPM-NJP-06}
\bibinfo{author}{L.~Columbo}, \bibinfo{author}{I.~M. Perrini},
  \bibinfo{author}{T.~Maggipinto}, \bibinfo{author}{M.~Brambilla},
\newblock \bibinfo{title}{3d self-organized patterns in the field profile of a
  semiconductor resonator},
\newblock \bibinfo{journal}{New Journal of Physics} \bibinfo{volume}{8}
  (\bibinfo{year}{2006}) \bibinfo{pages}{312}.
\bibitem[{Gopalakrishnan et~al.(2021)Gopalakrishnan, Panajotov, Taki, and
  Tlidi}]{GPT-PRL-21}
\bibinfo{author}{S.~S. Gopalakrishnan}, \bibinfo{author}{K.~Panajotov},
  \bibinfo{author}{M.~Taki}, \bibinfo{author}{M.~Tlidi},
\newblock \bibinfo{title}{Dissipative light bullets in kerr cavities:
  Multistability, clustering, and rogue waves},
\newblock \bibinfo{journal}{Phys. Rev. Lett.} \bibinfo{volume}{126}
  (\bibinfo{year}{2021}) \bibinfo{pages}{153902}.
  \DOIprefix\doi{10.1103/PhysRevLett.126.153902}.
\bibitem[{Sun et~al.(2023)Sun, Parra-Rivas, Mili\'an, Kartashov, Ferraro,
  Mangini, Jauberteau, Talenti, and Wabnitz}]{SPRM-PRL-23}
\bibinfo{author}{Y.~Sun}, \bibinfo{author}{P.~Parra-Rivas},
  \bibinfo{author}{C.~Mili\'an}, \bibinfo{author}{Y.~V. Kartashov},
  \bibinfo{author}{M.~Ferraro}, \bibinfo{author}{F.~Mangini},
  \bibinfo{author}{R.~Jauberteau}, \bibinfo{author}{F.~R. Talenti},
  \bibinfo{author}{S.~Wabnitz},
\newblock \bibinfo{title}{Robust three-dimensional high-order solitons and
  breathers in driven dissipative systems: A kerr cavity realization},
\newblock \bibinfo{journal}{Phys. Rev. Lett.} \bibinfo{volume}{131}
  (\bibinfo{year}{2023}) \bibinfo{pages}{137201}.
  \DOIprefix\doi{10.1103/PhysRevLett.131.137201}.
\bibitem[{Sun et~al.(2024)Sun, Parra-Rivas, Mangini, and Wabnitz}]{SPRM-CSF-24}
\bibinfo{author}{Y.~Sun}, \bibinfo{author}{P.~Parra-Rivas},
  \bibinfo{author}{F.~Mangini}, \bibinfo{author}{S.~Wabnitz},
\newblock \bibinfo{title}{Multidimensional localized states in externally
  driven kerr cavities with a parabolic spatiotemporal potential: A dimensional
  connection},
\newblock \bibinfo{journal}{Chaos, Solitons $\&$ Fractals}
  \bibinfo{volume}{183} (\bibinfo{year}{2024}) \bibinfo{pages}{114870}.
  \DOIprefix\doi{https://doi.org/10.1016/j.chaos.2024.114870}.
\bibitem[{Javaloyes(2016)}]{J_PRL_16}
\bibinfo{author}{J.~Javaloyes},
\newblock \bibinfo{title}{Cavity light bullets in passively mode-locked
  semiconductor lasers},
\newblock \bibinfo{journal}{Phys. Rev. Lett.} \bibinfo{volume}{116}
  (\bibinfo{year}{2016}) \bibinfo{pages}{043901}.
  \DOIprefix\doi{10.1103/PhysRevLett.116.043901}.
\bibitem[{Gurevich and Javaloyes(2017)}]{GJ-PRA-17}
\bibinfo{author}{S.~V. Gurevich}, \bibinfo{author}{J.~Javaloyes},
\newblock \bibinfo{title}{Spatial instabilities of light bullets in
  passively-mode-locked lasers},
\newblock \bibinfo{journal}{Phys. Rev. A} \bibinfo{volume}{96}
  (\bibinfo{year}{2017}) \bibinfo{pages}{023821}.
  \DOIprefix\doi{10.1103/PhysRevA.96.023821}.
\bibitem[{Pimenov et~al.(2018)Pimenov, Javaloyes, Gurevich, and
  Vladimirov}]{PJG-PTRA-18}
\bibinfo{author}{A.~Pimenov}, \bibinfo{author}{J.~Javaloyes},
  \bibinfo{author}{S.~V. Gurevich}, \bibinfo{author}{A.~G. Vladimirov},
\newblock \bibinfo{title}{Light bullets in a time-delay model of a
  wide-aperture mode-locked semiconductor laser},
\newblock \bibinfo{journal}{Philos. Trans. A. Math. Phys. Eng. Sci.}
  \bibinfo{volume}{376} (\bibinfo{year}{2018}).
  \DOIprefix\doi{10.1098/rsta.2017.0372}.
  \href{http://arxiv.org/abs/http://rsta.royalsocietypublishing.org/content/376/2124/20170372.full.pdf}{{\tt
  arXiv:http://rsta.royalsocietypublishing.org/content/376/2124/20170372.full.pdf}}.
\bibitem[{Dohmen et~al.(2020)Dohmen, Javaloyes, and Gurevich}]{DJG-Chaos-20}
\bibinfo{author}{F.~Dohmen}, \bibinfo{author}{J.~Javaloyes},
  \bibinfo{author}{S.~V. Gurevich},
\newblock \bibinfo{title}{Bound states of light bullets in passively
  mode-locked semiconductor lasers},
\newblock \bibinfo{journal}{Chaos: An Interdisciplinary Journal of Nonlinear
  Science} \bibinfo{volume}{30} (\bibinfo{year}{2020}) \bibinfo{pages}{063120}.
  \DOIprefix\doi{10.1063/5.0003227}.
  \href{http://arxiv.org/abs/https://doi.org/10.1063/5.0003227}{{\tt
  arXiv:https://doi.org/10.1063/5.0003227}}.
\bibitem[{Seidel et~al.(2020)Seidel, Perego, Javaloyes, and
  Gurevich}]{SPJG-Chaos-20}
\bibinfo{author}{T.~G. Seidel}, \bibinfo{author}{A.~M. Perego},
  \bibinfo{author}{J.~Javaloyes}, \bibinfo{author}{S.~V. Gurevich},
\newblock \bibinfo{title}{Discrete light bullets in passively mode-locked
  semiconductor lasers},
\newblock \bibinfo{journal}{Chaos} \bibinfo{volume}{30} (\bibinfo{year}{2020})
  \bibinfo{pages}{063102}. \DOIprefix\doi{10.1063/5.0002989}.
  \href{http://arxiv.org/abs/https://doi.org/10.1063/5.0002989}{{\tt
  arXiv:https://doi.org/10.1063/5.0002989}}.
\bibitem[{Fibich and Ilan(2004)}]{FI-OL-04}
\bibinfo{author}{G.~Fibich}, \bibinfo{author}{B.~Ilan},
\newblock \bibinfo{title}{Optical light bullets in a pure kerr medium},
\newblock \bibinfo{journal}{Opt. Lett.} \bibinfo{volume}{29}
  (\bibinfo{year}{2004}) \bibinfo{pages}{887--889}.
  \DOIprefix\doi{10.1364/OL.29.000887}.
\bibitem[{Parra-Rivas et~al.(2024)Parra-Rivas, Sun, Mangini, Ferraro, Zitelli,
  and Wabnitz}]{PRSM-PRA-24}
\bibinfo{author}{P.~Parra-Rivas}, \bibinfo{author}{Y.~Sun},
  \bibinfo{author}{F.~Mangini}, \bibinfo{author}{M.~Ferraro},
  \bibinfo{author}{M.~Zitelli}, \bibinfo{author}{S.~Wabnitz},
\newblock \bibinfo{title}{Pure quartic three-dimensional spatiotemporal kerr
  solitons in graded-index media},
\newblock \bibinfo{journal}{Phys. Rev. A} \bibinfo{volume}{109}
  (\bibinfo{year}{2024}) \bibinfo{pages}{033516}.
  \DOIprefix\doi{10.1103/PhysRevA.109.033516}.
\bibitem[{Tlidi and Gelens(2010)}]{TG-OL-10}
\bibinfo{author}{M.~Tlidi}, \bibinfo{author}{L.~Gelens},
\newblock \bibinfo{title}{High-order dispersion stabilizes dark dissipative
  solitons in all-fiber cavities},
\newblock \bibinfo{journal}{Opt. Lett.} \bibinfo{volume}{35}
  (\bibinfo{year}{2010}) \bibinfo{pages}{306--308}.
  \DOIprefix\doi{10.1364/OL.35.000306}.
\bibitem[{Parra-Rivas et~al.(2014)Parra-Rivas, Gomila, Leo, Coen, and
  Gelens}]{PGL-OL-14}
\bibinfo{author}{P.~Parra-Rivas}, \bibinfo{author}{D.~Gomila},
  \bibinfo{author}{F.~Leo}, \bibinfo{author}{S.~Coen},
  \bibinfo{author}{L.~Gelens},
\newblock \bibinfo{title}{Third-order chromatic dispersion stabilizes {K}err
  frequency combs},
\newblock \bibinfo{journal}{Opt. Lett.} \bibinfo{volume}{39}
  (\bibinfo{year}{2014}) \bibinfo{pages}{2971--2974}.
  \DOIprefix\doi{10.1364/OL.39.002971}.
\bibitem[{Parra-Rivas et~al.(2017)Parra-Rivas, Gomila, and Gelens}]{PGG_PRA_17}
\bibinfo{author}{P.~Parra-Rivas}, \bibinfo{author}{D.~Gomila},
  \bibinfo{author}{L.~Gelens},
\newblock \bibinfo{title}{Coexistence of stable dark- and bright-soliton {K}err
  combs in normal-dispersion resonators},
\newblock \bibinfo{journal}{Phys. Rev. A} \bibinfo{volume}{95}
  (\bibinfo{year}{2017}) \bibinfo{pages}{053863}.
  \DOIprefix\doi{10.1103/PhysRevA.95.053863}.
\bibitem[{Tam et~al.(2020)Tam, Alexander, Blanco-Redondo, and
  de~Sterke}]{TABB-PRA-20}
\bibinfo{author}{K.~K.~K. Tam}, \bibinfo{author}{T.~J. Alexander},
  \bibinfo{author}{A.~Blanco-Redondo}, \bibinfo{author}{C.~M. de~Sterke},
\newblock \bibinfo{title}{Generalized dispersion {K}err solitons},
\newblock \bibinfo{journal}{Phys. Rev. A} \bibinfo{volume}{101}
  (\bibinfo{year}{2020}) \bibinfo{pages}{043822}.
  \DOIprefix\doi{10.1103/PhysRevA.101.043822}.
\bibitem[{Karlsson and Höök(1994)}]{KH-OC-94}
\bibinfo{author}{M.~Karlsson}, \bibinfo{author}{A.~Höök},
\newblock \bibinfo{title}{Soliton-like pulses governed by fourth order
  dispersion in optical fibers},
\newblock \bibinfo{journal}{Opt. Commun.} \bibinfo{volume}{104}
  (\bibinfo{year}{1994}) \bibinfo{pages}{303}.
\bibitem[{Blanco-Redondo et~al.(2016)Blanco-Redondo, de~Sterke, Sipe, Krauss,
  Eggleton, and Husko}]{BSS-NAC-16}
\bibinfo{author}{A.~Blanco-Redondo}, \bibinfo{author}{C.~M. de~Sterke},
  \bibinfo{author}{J.~E. Sipe}, \bibinfo{author}{T.~F. Krauss},
  \bibinfo{author}{B.~J. Eggleton}, \bibinfo{author}{C.~Husko},
\newblock \bibinfo{title}{Pure-quartic solitons},
\newblock \bibinfo{journal}{Nature Communications} \bibinfo{volume}{7}
  (\bibinfo{year}{2016}) \bibinfo{pages}{10427}.
  \DOIprefix\doi{10.1038/ncomms10427}.
\bibitem[{de~Sterke et~al.(2021)de~Sterke, Runge, Hudson, and
  Blanco-Redondo}]{DRH-APLP-21}
\bibinfo{author}{C.~M. de~Sterke}, \bibinfo{author}{A.~F.~J. Runge},
  \bibinfo{author}{D.~D. Hudson}, \bibinfo{author}{A.~Blanco-Redondo},
\newblock \bibinfo{title}{Pure-quartic solitons and their
  generalizations—theory and experiments},
\newblock \bibinfo{journal}{APL Photonics} \bibinfo{volume}{6}
  (\bibinfo{year}{2021}) \bibinfo{pages}{091101}.
  \DOIprefix\doi{10.1063/5.0059525}.
  \href{http://arxiv.org/abs/https://doi.org/10.1063/5.0059525}{{\tt
  arXiv:https://doi.org/10.1063/5.0059525}}.
\bibitem[{Tlidi et~al.(2007)Tlidi, Mussot, Louvergneaux, Kozyreff, Vladimirov,
  and Taki}]{TML-OL-07}
\bibinfo{author}{M.~Tlidi}, \bibinfo{author}{A.~Mussot},
  \bibinfo{author}{E.~Louvergneaux}, \bibinfo{author}{G.~Kozyreff},
  \bibinfo{author}{A.~G. Vladimirov}, \bibinfo{author}{M.~Taki},
\newblock \bibinfo{title}{Control and removal of modulational instabilities in
  low-dispersion photonic crystal fiber cavities},
\newblock \bibinfo{journal}{Opt. Lett.} \bibinfo{volume}{32}
  (\bibinfo{year}{2007}) \bibinfo{pages}{662--664}.
  \DOIprefix\doi{10.1364/OL.32.000662}.
\bibitem[{Droques et~al.(2013)Droques, Kudlinski, Bouwmans, Martinelli, Mussot,
  Armaroli, and Biancalana}]{DKB-OL-13}
\bibinfo{author}{M.~Droques}, \bibinfo{author}{A.~Kudlinski},
  \bibinfo{author}{G.~Bouwmans}, \bibinfo{author}{G.~Martinelli},
  \bibinfo{author}{A.~Mussot}, \bibinfo{author}{A.~Armaroli},
  \bibinfo{author}{F.~Biancalana},
\newblock \bibinfo{title}{Fourth-order dispersion mediated modulation
  instability in dispersion oscillating fibers},
\newblock \bibinfo{journal}{Opt. Lett.} \bibinfo{volume}{38}
  (\bibinfo{year}{2013}) \bibinfo{pages}{3464--3467}.
  \DOIprefix\doi{10.1364/OL.38.003464}.
\bibitem[{Lega et~al.(1994)Lega, Moloney, and Newell}]{Lega_1994}
\bibinfo{author}{J.~Lega}, \bibinfo{author}{J.~Moloney},
  \bibinfo{author}{A.~Newell},
\newblock \bibinfo{title}{Swift-hohenberg equation for lasers},
\newblock \bibinfo{journal}{Phys. Rev. Lett.} \bibinfo{volume}{73}
  (\bibinfo{year}{1994}) \bibinfo{pages}{2978--2981}.
\bibitem[{Khanolkar et~al.(2021)Khanolkar, Zang, and Chong}]{KZC-PhotRes-21}
\bibinfo{author}{A.~Khanolkar}, \bibinfo{author}{Y.~Zang},
  \bibinfo{author}{A.~Chong},
\newblock \bibinfo{title}{Complex {S}wift-{H}ohenberg equation dissipative
  soliton fiber laser},
\newblock \bibinfo{journal}{Photon. Res.} \bibinfo{volume}{9}
  (\bibinfo{year}{2021}) \bibinfo{pages}{1033--1038}.
  \DOIprefix\doi{10.1364/PRJ.419686}.
\bibitem[{Huyet et~al.(1995)Huyet, Martinoni, Tredicce, and Rica}]{HMT-PRL-95}
\bibinfo{author}{G.~Huyet}, \bibinfo{author}{M.~C. Martinoni},
  \bibinfo{author}{J.~R. Tredicce}, \bibinfo{author}{S.~Rica},
\newblock \bibinfo{title}{Spatiotemporal dynamics of lasers with a large
  {F}resnel number},
\newblock \bibinfo{journal}{Phys. Rev. Lett.} \bibinfo{volume}{75}
  (\bibinfo{year}{1995}) \bibinfo{pages}{4027--4030}.
  \DOIprefix\doi{10.1103/PhysRevLett.75.4027}.
\bibitem[{Stali\ifmmode~\bar{u}\else \={u}\fi{}nas
  et~al.(1997)Stali\ifmmode~\bar{u}\else \={u}\fi{}nas, \ifmmode~\check{S}\else
  \v{S}\fi{}lekys, and Weiss}]{SSW-PRL-97}
\bibinfo{author}{K.~Stali\ifmmode~\bar{u}\else \={u}\fi{}nas},
  \bibinfo{author}{G.~\ifmmode~\check{S}\else \v{S}\fi{}lekys},
  \bibinfo{author}{C.~O. Weiss},
\newblock \bibinfo{title}{Nonlinear pattern formation in active optical
  systems: Shocks, domains of tilted waves, and cross-roll patterns},
\newblock \bibinfo{journal}{Phys. Rev. Lett.} \bibinfo{volume}{79}
  (\bibinfo{year}{1997}) \bibinfo{pages}{2658--2661}.
  \DOIprefix\doi{10.1103/PhysRevLett.79.2658}.
\bibitem[{Mercier and Moloney(2002)}]{MM-PRE-02}
\bibinfo{author}{J.-F. Mercier}, \bibinfo{author}{J.~V. Moloney},
\newblock \bibinfo{title}{Derivation of semiconductor laser mean-field and
  {S}wift-{H}ohenberg equations},
\newblock \bibinfo{journal}{Phys. Rev. E} \bibinfo{volume}{66}
  (\bibinfo{year}{2002}) \bibinfo{pages}{036221}.
  \DOIprefix\doi{10.1103/PhysRevE.66.036221}.
\bibitem[{Gelens et~al.(2007)Gelens, Van~der Sande, Tassin, Tlidi, Kockaert,
  Gomila, Veretennicoff, and Danckaert}]{GCdST-PRA-07}
\bibinfo{author}{L.~Gelens}, \bibinfo{author}{G.~Van~der Sande},
  \bibinfo{author}{P.~Tassin}, \bibinfo{author}{M.~Tlidi},
  \bibinfo{author}{P.~Kockaert}, \bibinfo{author}{D.~Gomila},
  \bibinfo{author}{I.~Veretennicoff}, \bibinfo{author}{J.~Danckaert},
\newblock \bibinfo{title}{Impact of nonlocal interactions in dissipative
  systems: Towards minimal-sized localized structures},
\newblock \bibinfo{journal}{Phys. Rev. A} \bibinfo{volume}{75}
  (\bibinfo{year}{2007}) \bibinfo{pages}{063812}.
  \DOIprefix\doi{10.1103/PhysRevA.75.063812}.
\bibitem[{Kockaert et~al.(2009)Kockaert, Tassin, Veretennicoff, der Sande, and
  Tlidi}]{KTV-JOSAB-09}
\bibinfo{author}{P.~Kockaert}, \bibinfo{author}{P.~Tassin},
  \bibinfo{author}{I.~Veretennicoff}, \bibinfo{author}{G.~V. der Sande},
  \bibinfo{author}{M.~Tlidi},
\newblock \bibinfo{title}{Beyond the zero-diffraction regime in optical
  cavities with a left-handed material},
\newblock \bibinfo{journal}{J. Opt. Soc. Am. B} \bibinfo{volume}{26}
  (\bibinfo{year}{2009}) \bibinfo{pages}{B148--B155}.
  \DOIprefix\doi{10.1364/JOSAB.26.00B148}.
\bibitem[{Tlidi et~al.(2011)Tlidi, Kockaert, and Gelens}]{TKG-PRA-11}
\bibinfo{author}{M.~Tlidi}, \bibinfo{author}{P.~Kockaert},
  \bibinfo{author}{L.~Gelens},
\newblock \bibinfo{title}{Dark localized structures in a cavity filled with a
  left-handed material},
\newblock \bibinfo{journal}{Phys. Rev. A} \bibinfo{volume}{84}
  (\bibinfo{year}{2011}) \bibinfo{pages}{013807}.
  \DOIprefix\doi{10.1103/PhysRevA.84.013807}.
\bibitem[{Gurevich et~al.(2024)Gurevich, Maucher, and Javaloyes}]{GMJ-PRR-24}
\bibinfo{author}{S.~V. Gurevich}, \bibinfo{author}{F.~Maucher},
  \bibinfo{author}{J.~Javaloyes},
\newblock \bibinfo{title}{Aberration-driven tilted emission in degenerate
  cavities},
\newblock \bibinfo{journal}{Phys. Rev. Res.} \bibinfo{volume}{6}
  (\bibinfo{year}{2024}) \bibinfo{pages}{013166}.
  \DOIprefix\doi{10.1103/PhysRevResearch.6.013166}.
\bibitem[{Bartolo et~al.(2022)Bartolo, Vigne, Marconi, Beaudoin, Pantzas,
  Sagnes, Huyet, Maucher, Gurevich, Javaloyes, Garnache, and
  Giudici}]{BVM-Optica-22}
\bibinfo{author}{A.~Bartolo}, \bibinfo{author}{N.~Vigne},
  \bibinfo{author}{M.~Marconi}, \bibinfo{author}{G.~Beaudoin},
  \bibinfo{author}{K.~Pantzas}, \bibinfo{author}{I.~Sagnes},
  \bibinfo{author}{G.~Huyet}, \bibinfo{author}{F.~Maucher},
  \bibinfo{author}{S.~V. Gurevich}, \bibinfo{author}{J.~Javaloyes},
  \bibinfo{author}{A.~Garnache}, \bibinfo{author}{M.~Giudici},
\newblock \bibinfo{title}{Temporal localized {T}uring patterns in mode-locked
  semiconductor lasers},
\newblock \bibinfo{journal}{Optica} \bibinfo{volume}{9} (\bibinfo{year}{2022})
  \bibinfo{pages}{1386--1393}. \DOIprefix\doi{10.1364/OPTICA.471006}.
\bibitem[{Bartolo et~al.(2023)Bartolo, Vigne, Marconi, Beaudoin, Gratiet,
  Pantzas, Sagnes, Garnache, and Giudici}]{BVM-PR-23}
\bibinfo{author}{A.~Bartolo}, \bibinfo{author}{N.~Vigne},
  \bibinfo{author}{M.~Marconi}, \bibinfo{author}{G.~Beaudoin},
  \bibinfo{author}{L.~L. Gratiet}, \bibinfo{author}{K.~Pantzas},
  \bibinfo{author}{I.~Sagnes}, \bibinfo{author}{A.~Garnache},
  \bibinfo{author}{M.~Giudici},
\newblock \bibinfo{title}{Spatiotemporally reconfigurable light in degenerate
  laser cavities},
\newblock \bibinfo{journal}{Photon. Res.} \bibinfo{volume}{11}
  (\bibinfo{year}{2023}) \bibinfo{pages}{1751--1756}.
  \DOIprefix\doi{10.1364/PRJ.495892}.
\bibitem[{Schelte et~al.(2019)Schelte, Pimenov, Vladimirov, Javaloyes, and
  Gurevich}]{SPV-OL-19}
\bibinfo{author}{C.~Schelte}, \bibinfo{author}{A.~Pimenov},
  \bibinfo{author}{A.~G. Vladimirov}, \bibinfo{author}{J.~Javaloyes},
  \bibinfo{author}{S.~V. Gurevich},
\newblock \bibinfo{title}{Tunable {K}err frequency combs and temporal localized
  states in time-delayed {G}ires-{T}ournois interferometers},
\newblock \bibinfo{journal}{Opt. Lett.} \bibinfo{volume}{44}
  (\bibinfo{year}{2019}) \bibinfo{pages}{4925--4928}.
  \DOIprefix\doi{10.1364/OL.44.004925}.
\bibitem[{Seidel et~al.(2022)Seidel, Javaloyes, and Gurevich}]{SJG-OL-22}
\bibinfo{author}{T.~G. Seidel}, \bibinfo{author}{J.~Javaloyes},
  \bibinfo{author}{S.~V. Gurevich},
\newblock \bibinfo{title}{A normal form for frequency combs and localized
  states in kerr--gires--tournois interferometers},
\newblock \bibinfo{journal}{Opt. Lett.} \bibinfo{volume}{47}
  (\bibinfo{year}{2022}) \bibinfo{pages}{2979--2982}.
  \DOIprefix\doi{10.1364/OL.457777}.
\bibitem[{Koch et~al.(2022)Koch, Seidel, Gurevich, and Javaloyes}]{KSG-OL-22}
\bibinfo{author}{E.~Koch}, \bibinfo{author}{T.~Seidel},
  \bibinfo{author}{S.~Gurevich}, \bibinfo{author}{J.~Javaloyes},
\newblock \bibinfo{title}{Square-wave generation in vertical external-cavity
  kerr-gires-tournois interferometers},
\newblock \bibinfo{journal}{Optics Letters} \bibinfo{volume}{47}
  (\bibinfo{year}{2022}) \bibinfo{pages}{4343--4346}.
  \DOIprefix\doi{10.1364/OL.468236}.
\bibitem[{St{\"o}hr et~al.(2023)St{\"o}hr, Koch, Javaloyes, Gurevich, and
  Wolfrum}]{SKJGW_Chaos_23}
\bibinfo{author}{M.~St{\"o}hr}, \bibinfo{author}{E.~R. Koch},
  \bibinfo{author}{J.~Javaloyes}, \bibinfo{author}{S.~V. Gurevich},
  \bibinfo{author}{M.~Wolfrum},
\newblock \bibinfo{title}{Square waves and {B}ykov {T}-points in a delay
  algebraic model for the {K}err--{G}ires--{T}ournois interferometer},
\newblock \bibinfo{journal}{Chaos} \bibinfo{volume}{33} (\bibinfo{year}{2023}).
  \DOIprefix\doi{10.1063/5.0173320}.
\bibitem[{Seidel et~al.(2024)Seidel, Javaloyes, and Gurevich}]{SJG-OL-24}
\bibinfo{author}{T.~G. Seidel}, \bibinfo{author}{J.~Javaloyes},
  \bibinfo{author}{S.~V. Gurevich},
\newblock \bibinfo{title}{Normal dispersion kerr cavity solitons: beyond the
  mean-field limit},
\newblock \bibinfo{journal}{Opt. Lett.} \bibinfo{volume}{49}
  (\bibinfo{year}{2024}) \bibinfo{pages}{7008--7011}.
  \DOIprefix\doi{10.1364/OL.538135}.
\bibitem[{Koch et~al.(2026)Koch, Greve, Javaloyes, and Gurevich}]{KGJ-CSF-26}
\bibinfo{author}{E.~R. Koch}, \bibinfo{author}{D.~Greve},
  \bibinfo{author}{J.~Javaloyes}, \bibinfo{author}{S.~V. Gurevich},
\newblock \bibinfo{title}{Thermo-optical dynamics in injected {K}err
  micro-cavities with strong time-delayed feedback},
\newblock \bibinfo{journal}{Chaos, Solitons $\&$ Fractals}
  \bibinfo{volume}{202} (\bibinfo{year}{2026}) \bibinfo{pages}{117595}.
  \DOIprefix\doi{https://doi.org/10.1016/j.chaos.2025.117595}.
\bibitem[{Hunkemoeller et~al.(2026)Hunkemoeller, Seidel, Javaloyes, and
  Gurevich}]{HSJ-Chaos-26}
\bibinfo{author}{M.~Hunkemoeller}, \bibinfo{author}{T.~G. Seidel},
  \bibinfo{author}{J.~Javaloyes}, \bibinfo{author}{S.~V. Gurevich},
\newblock \bibinfo{title}{Impact of phase modulation on the dynamics of
  temporal localized structures in injected {K}err microcavities},
\newblock \bibinfo{journal}{Chaos: An Interdisciplinary Journal of Nonlinear
  Science} \bibinfo{volume}{36} (\bibinfo{year}{2026}) \bibinfo{pages}{073136}.
  \DOIprefix\doi{10.1063/5.0332806}.
\bibitem[{Seidel et~al.(2022)Seidel, Gurevich, and Javaloyes}]{SGJ-PRL-22}
\bibinfo{author}{T.~G. Seidel}, \bibinfo{author}{S.~V. Gurevich},
  \bibinfo{author}{J.~Javaloyes},
\newblock \bibinfo{title}{Conservative solitons and reversibility in time
  delayed systems},
\newblock \bibinfo{journal}{Phys. Rev. Lett.} \bibinfo{volume}{128}
  (\bibinfo{year}{2022}) \bibinfo{pages}{083901}.
  \DOIprefix\doi{10.1103/PhysRevLett.128.083901}.
\bibitem[{Lugiato and Lefever(1987)}]{LL-PRL-87}
\bibinfo{author}{L.~A. Lugiato}, \bibinfo{author}{R.~Lefever},
\newblock \bibinfo{title}{Spatial dissipative structures in passive optical
  systems},
\newblock \bibinfo{journal}{Phys. Rev. Lett.} \bibinfo{volume}{58}
  (\bibinfo{year}{1987}) \bibinfo{pages}{2209--2211}.
  \DOIprefix\doi{10.1103/PhysRevLett.58.2209}.
\bibitem[{Haelterman et~al.(1992)Haelterman, Trillo, and
  Wabnitz}]{HTW-OptCom-92}
\bibinfo{author}{M.~Haelterman}, \bibinfo{author}{S.~Trillo},
  \bibinfo{author}{S.~Wabnitz},
\newblock \bibinfo{title}{Dissipative modulation instability in a nonlinear
  dispersive ring cavity},
\newblock \bibinfo{journal}{Optics Communications} \bibinfo{volume}{91}
  (\bibinfo{year}{1992}) \bibinfo{pages}{401--407}.
  \DOIprefix\doi{https://doi.org/10.1016/0030-4018(92)90367-Z}.
\bibitem[{Siegman(1986)}]{SIEGMAN-BOOK}
\bibinfo{author}{A.~E. Siegman}, \bibinfo{title}{Lasers},
  \bibinfo{publisher}{University Science Books, Mill Valley, CA},
  \bibinfo{year}{1986}.
\bibitem[{Uecker et~al.(2014)Uecker, Wetzel, and Rademacher}]{uecker2014}
\bibinfo{author}{H.~Uecker}, \bibinfo{author}{D.~Wetzel},
  \bibinfo{author}{J.~D.~M. Rademacher},
\newblock \bibinfo{title}{pde2path - a matlab package for continuation and
  bifurcation in 2d elliptic systems},
\newblock \bibinfo{journal}{Num. Math.: Theory, Methods and App.}
  \bibinfo{volume}{7} (\bibinfo{year}{2014}) \bibinfo{pages}{58--106}.
  \DOIprefix\doi{10.1017/S1004897900000295}.

\end{thebibliography}
\end{document}